\documentclass[12pt,cite,epsf,epsfig,psfrag]{article}
\usepackage{epsfig}
\usepackage{amsmath, graphics, setspace}
\usepackage{color}
\usepackage{graphicx}
\usepackage{psfrag}
\usepackage{amssymb}
\usepackage{subfigure}

\newcommand{\be}{\begin{equation}}
\newcommand{\ee}{\end{equation}}
\newcommand{\bea}{\begin{eqnarray}}
\newcommand{\eea}{\end{eqnarray}}

\newcommand{\mathsym}[1]{{}}
\newcommand{\unicode}[1]{{}}
\begin{document}
\topmargin 0pt
\oddsidemargin 0mm

\vspace{2mm}

\begin{center}

{\Large \bf {Schwinger Effect and Negative Differential Conductivity in  Holographic Models}}

\vspace{1 cm}

{Shankhadeep Chakrabortty$^{\dagger,*}$\footnote{E-mail: shankha@imsc.res.in, shankhadeep.chakrabortty@iiserpune.ac.in} and B.Sathiapalan$^*$\footnote{E-mail: bala@imsc.res.in}}

 \vspace{4mm}

{\em
$^{\dagger}$Indian Institute of Science Education and Research, Pune, India 411008.}

{\em
$^{*}$Institute of Mathematical Sciences, Taramani, Chennai, India 600113.}

\end{center}

\vspace{1 cm}

\begin{abstract}
The consequences of the Schwinger effect for conductivity is computed for strong coupling systems using holography. 
The one loop diagram on the flavor brane introduces
an $O({\lambda \over N_c})$ imaginary part in the effective action for a Maxwell flavor gauge field.  
This in turn introduces a real conductivity in an otherwise insulating phase of the boundary theory. 
Moreover in certain regions of parameter space the differential conductivity is {\em negative}. 
This is computed in the context of the Sakai-Sugimoto model.
\end{abstract}

\newpage
\tableofcontents
\section{Introduction}

The phenomenon of electron positron pair production in the presence of an electric field was described quantitatively in the classic work of 
 Schwinger \cite{Schwinger:1951nm}. He wrote down an expression for the probability of particle production 
(per unit space time volume) by solving the Dirac equation  in a  uniform and constant background 
electric field $E$ and obtaining the electron Green function. The expression is 
\begin{eqnarray}
P (E,m) = 1-e^{-\Gamma V T}.  
\end{eqnarray}
with
\begin{eqnarray}
\Gamma = {E^2\over 4\pi^3}\sum_{n=1}^{\infty} {1\over n^2}e^{-\pi m^2 n\over |E|}.
\end{eqnarray}

A noteworthy feature is that there is an exponential damping with the mass of the electron which signifies that it is a tunneling 
phenomenon. Thus for laboratory electric fields the magnitude is negligible and unobservable. 

Schwinger's derivation treats the electric field as classical and also ignores higher order effects such as the Coulomb force between the electrons. As shown in \cite{Semenoff} the presence of 
the Coulomb potential modifies the solution for large values of the electric field. It is found that for $E>E_c$, (critical field), there is no potential barrier and hence no exponential suppression. 
In this approximation the potential has the form  $ 2m-Ed-{\alpha\over d}$ \cite{Semenoff}, where $\alpha$ is proportional to the electric charge.
  When 
\begin{eqnarray}
 	\label{Ecrit}
E=E_c={m^2\over \alpha},
\end{eqnarray}

the maximum of the potential is zero and there is no barrier. 

It is clear that this is non perturbative in $\alpha$. One can then pose the same question in ${\cal N} =4$ Super 
Yang-Mills theory which is also in the Coulomb phase. Here in the planar strong coupling limit one can resort to 
the AdS/CFT correspondence and get an exact answer.
Again one finds a critical electrical field \cite{Semenoff}. 

This is also what one expects from string theory in flat space: An open string in an electric field is a quark-antiquark pair 
in an electric field. When the force due to the electric field is stronger than the string tension, 
the effective string tension becomes zero and one can expect unsuppressed production of quark antiquark pairs from the vacuum. 
The signal of this is that the Dirac-Born-Infeld action becomes zero (and starts to become imaginary) at the critical electric 
field.

One can ask whether this effect can be observed in condensed matter systems by its consequences on conductivity \cite{Sondhi}. 
The effect of Schwinger pair production and its influences on conductivity was shown using 
holographic calculations in \cite{Karch-metallic, Karch-Sondhi}. Probe branes representing flavor quarks were introduced in an 
AdS-Schwarzschild  background and placed in an electric field and the conductivity 
was calculated \cite{Karch1}. 
Similar systems were also studied in \cite{Myers,Bergmann,Kundu}. The electric field induced conductivity was found to be a 
non-linear function of the electric field, and more importantly the critical 
field turns out to be zero\cite{Bergmann}. In the  bulk calculation the absence of a critical field can be traced to the warp 
factors. For an arbitrarily small electric field the warp factors reduce the effective tension of 
the string and at some value of the radial coordinate
$u=u^*$, the DBI action threatens to turn complex. It can be shown then that this unphysical behavior is resolved if one 
assumes that a non zero current is induced. This current is present even at 
zero temperature and can be attributed to the Schwinger effect without, however, any exponential suppression. In the boundary 
theory this can be attributed to the strong coupling effects which reduce the 
critical field to zero.\footnote{Naively, if one lets $\alpha \to \infty$ in (\ref{Ecrit}), $E_c\to 0$.}

 It has also been shown that this effect can be described in terms of an effective event horizon at $u=u^*$, 
not in the bulk metric, but in the open string metric on the brane \cite{Kim,Sonner}. In fact many of the results derived for
black hole backgrounds, such as the proof of the fluctuation dissipation theorem \cite{Son-Teaney} can be shown to be 
directly applicable in this situation also. 

There are situations, such as the one described in the Sakai-Sugimoto model \cite{SakSug}, where the branes do not reach as 
far as $u^*$. In this case this effect is not there and the material 
is insulating even when there is an electric field 
\cite{Bergmann}. For the boundary theory this implies that the quarks are too heavy and the strong coupling interactions
are not sufficient to remove the barrier. 

In the case that there is no barrier (due to strong coupling effects), the pair production takes place ``classically'' i.e. no
quantum mechanical tunneling is involved. One doesn't really 
need the Schwinger formula to calculate conductivity. This is the situation dealt with in the classical bulk calculations 
of \cite{Karch-metallic,Sonner, Kundu}. However one can ask the question whether, 
in the case where there is a barrier, as in the Sakai-Sugimoto model, where chiral symmetry breaking induces a quark mass 
and puts it in an insulating phase \cite{Kundu}, 
one can really use the Schwinger formula to calculate the conductivity. Schwinger's calculation is a one loop calculation 
and since the quarks circulating in the loop transform in the fundamental 
of the color group rather than the adjoint, this is a $1/N$ effect - beyond the planar approximation. If one includes the 
Schwinger correction (exponentially suppressed) to the Maxwell action 
(as calculated in \cite{Urrutia}) this would induce 
corrections to conductivity that is proportional to the electric field.

In this paper we address this question. The Maxwell action on the brane (which is the leading term obtained in expanding the 
DBI action)
is modified by the addition of the one loop effective action, which in the presence of a background electric field has an 
imaginary part explicitly. If one calculates conductivity using the Kubo formula, 
from the Green's function, which can be evaluated by the usual AdS/CFT
prescription, one expects a non zero real part. Indeed this is what we find. Thus we can conclude that what was thought to
be an insulating phase actually has a small conductivity.

Whether this is large enough to be observable experimentally depends crucially on the mass of the fermion. In metals the 
role of particle and anti particle is played by 
fermionic excitations at the Fermi surface which is ungapped and one can expect such effects. In ordinary metals the dispersion 
relation is non relativistic and the Schwinger formula would not be directly applicable although. Moreover external electric 
fields are screened in metals. It has been suggested that  systems such as graphene one can look for this effect \cite{ACG}. 
In graphene the electrons and holes are effectively massless at the "Dirac points" and obey a relativistic dispersion relation. Nevertheless
the density of states is small enough that screening effect is not strong.
Also the rate may be large enough to be measurable \cite{ACG}. The holographic calculation would then be applicable in a strong 
coupling version of this.

Another interesting feature that arises in this calculation is that there are regions where the real part of the conductivity is 
negative.
While this is unphysical ordinarily and would be a sign that the approximations are breaking down, this need not be so when there is 
a background electric field, from which energy can be pumped into the system. Thus in semiconductor physics it is well known that,
 Gunn diodes and tunnel diodes display negative differential resistivity in the presence of large external fields. A holographic 
calculation showing this phenomena has been done recently \cite{Nakamura1, Nakamura2, Nakamura-Ooguri}. Our main point of departure 
is that we are 
working in a region where the phenomenon is due to a one loop effect in the flavor brane in the bulk.

In supersymmetric BPS configurations of branes, the low energy action is fixed by the symmetries. However when supersymmetry is 
spontaneously broken by finite temperature effects one can expect finite loop corrections. Loop corrections can involve open 
strings with ($\mathcal{A}$) both ends on the flavor brane or ($\mathcal{B}$) 
with one end on the flavor brane and one end on the colored brane. 

\begin{figure}[h]
 \centering
 \mbox{\subfigure[]{\includegraphics[width=3.5 cm]{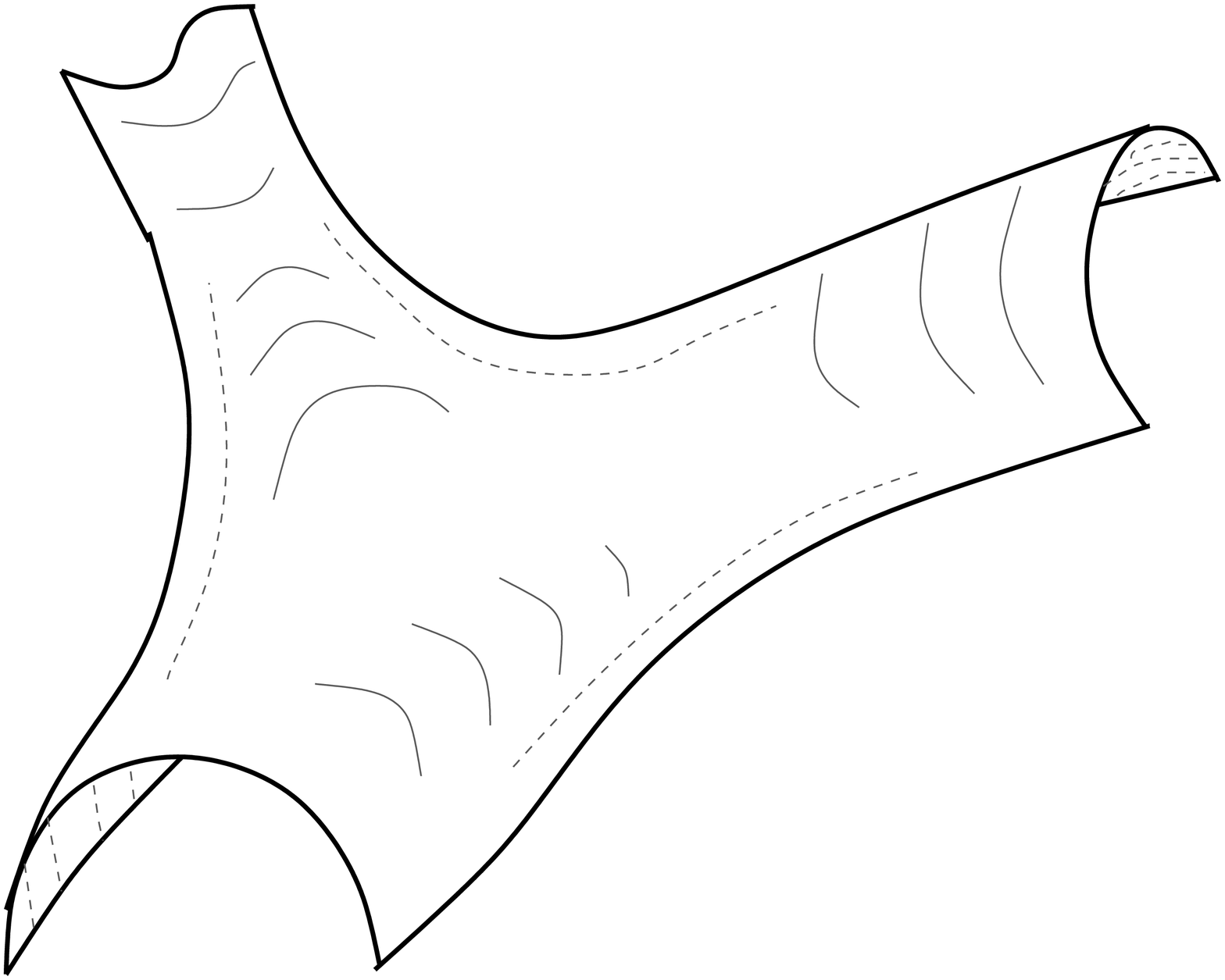}}
 \quad
 \subfigure[]{\includegraphics[width=3.5 cm]{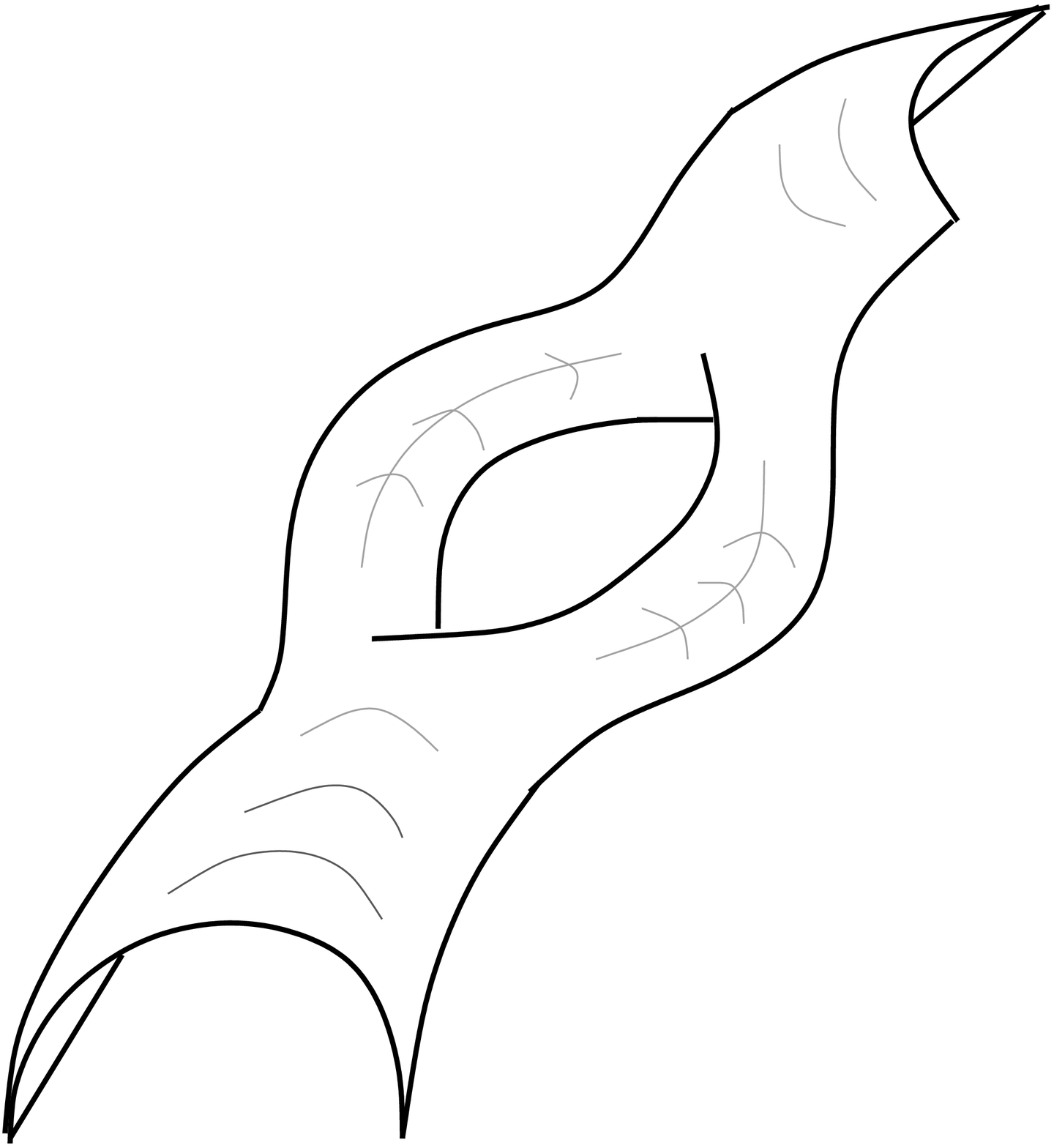} }
  \quad
 \subfigure[]{\includegraphics[width=4.5 cm]{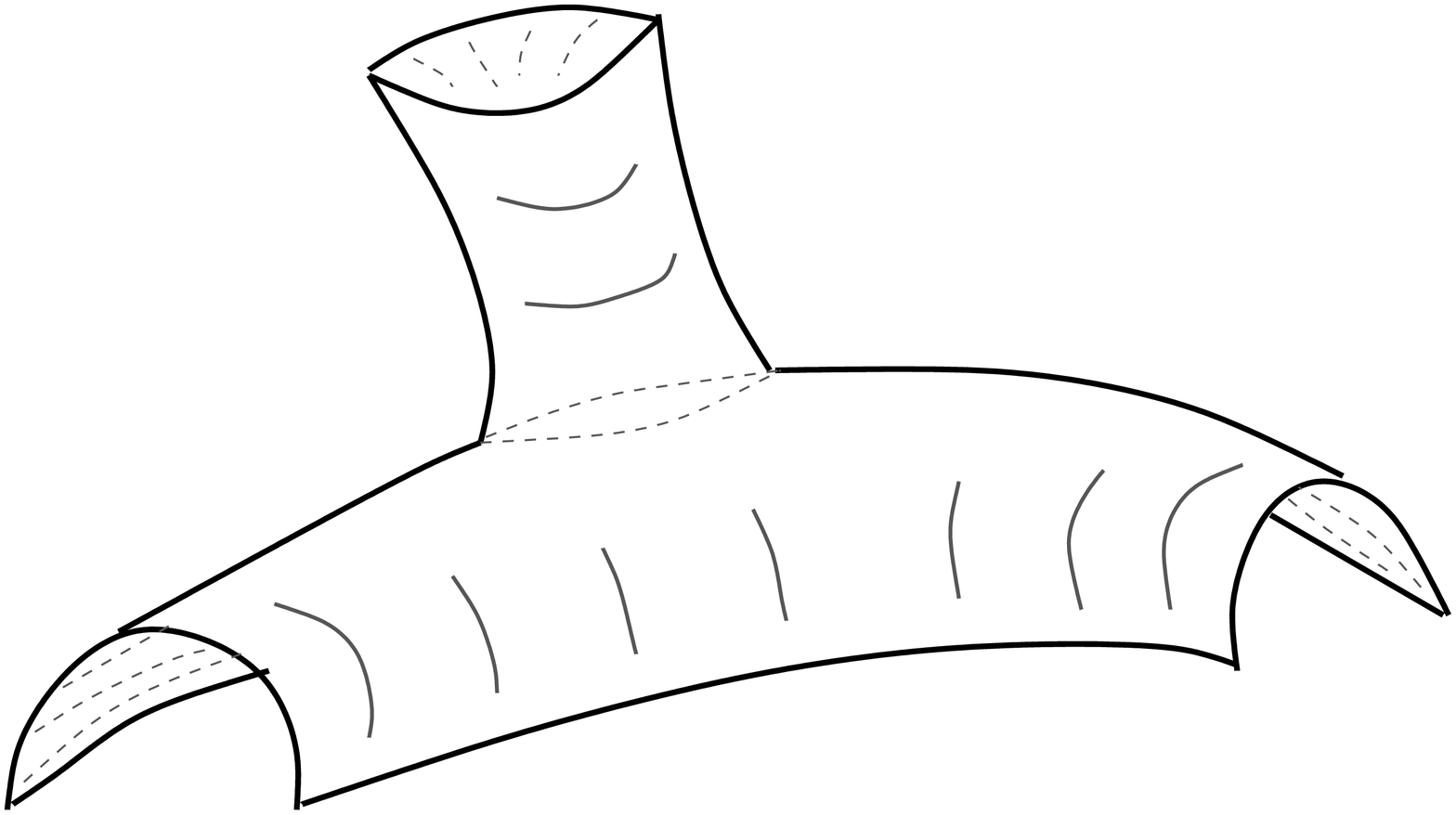} }}
 \caption{Figure (a) and (b) describe the tree level and the one loop diagrams (type $\mathcal{A}$) respectively for open string. 
Figure (c) shows a
 a one loop diagram with one end of the open string on the colored brane (type $\mathcal{B}$). This dualizes to a closed string tree diagram.}
 \label{embed1}
 \end{figure}

In case ($\mathcal{B}$) the loop diagram
with an infinite number of massive modes running around the loop dualizes to a tree diagram involving closed string modes 
connecting the flavor and color branes (see figure \ref{embed1}). One can see that they are of $O(\lambda)$. 
These are the same class of diagrams that generate the AdS background in the first place. So the question arise as to 
whether we are double counting. Are these diagrams (of type ($\mathcal{B}$)) to be considered or are they already included when the 
flavor brane is placed in an AdS space? In general it is the the UV limit of the open string loop that is reproduced by the 
graviton tree diagram. The finite part of the open string loop requires all the massive closed string tree diagrams. 
This continues to be true in the decoupling limit taken for the AdS/CFT correspondence. On the open string side, on the brane, 
in the $\alpha '\to 0$ limit  only the massless open string particles traverse the loop. However the closed string side still requires 
the full string theory  and the massive modes also contribute and their effect survives in this limit \cite{SarkSath}.\footnote{ In some situations the massive mode 
contributions cancel and the duality holds for the massless modes on both sides\cite{SarkSath}.}  Thus one is tempted to conclude that 
one has to explicitly calculate the finite part of a gauge theory loop diagram (with colored particle in the loop) on the brane placed in
 an AdS background. 
The finite contribution  in the presence of an electric field has an imaginary part corresponding to Schwinger
pair production.  In situations where the open string metric develops a horizon  such effects are seen, in the gravity dual, 
already at the tree level on the brane. But in general it is hard to imagine how an imaginary part can develop in an effective action except
 through loop effects. We leave the issue of type ($\mathcal{B}$) diagrams as an open question for the moment. 

  However 
diagrams of type ($\mathcal{A}$) are not included in the dual picture and must be computed - they are of order $\lambda  \times {1\over N_c}\times N_f$.  As an example if we 
have two flavor branes separated by a small amount, we have $SU(2)$ broken to $U(1)$. The effective action for the $U(1)$ photon 
will get corrected by the loop with $W^\pm$ or its fermionic partners running around.  This kind of symmetry  breaking occurs for 
instance in the holographic description of strong coupling BCS theory \cite{KSSS}. 

The actual calculation of these classes of diagrams is not very different - they differ only by an overall numerical factor which is of $O(N_c/N_f)$. Furthermore while the original Schwinger calculation involved only fermions in the loop in the supersymmetric theory there are fermions as well as bosons. The expressions for arbitrary spin are available in the literature \cite{Semenoff}. In this paper we are interested in probing the qualitative effects of the Schwinger effect, so we restrict ourselves to the fermionic contribution. 

Having discussed an overview and the motivation for our holographic study of Schwinger pair production in strongly coupled system, 
here we briefly mention about the specific strategies we follow in the subsequent sections. Our aim is to 
capture the effect of pair production process on the electrical conductivity in the context of Sakai-Sugimoto model. 
In section (\ref{one}), we briefly discuss the holographic aspects of Sakai-Sugimoto model. 
The bulk gravity theory of Sakai-Sugimoto model is built on the consideration of $N_f$ number of flavor $D8/\bar{D8}$ branes 
probing the near horizon geometry of type IIA supergravity background corresponding to the $N_c$ 
number of color $D4$ branes. Using the 
$D4-D8-\bar{D8}$ brane picture, we describe how the model holographically allows a deconfined boundary phase with broken
chiral symmetry. 
In section (\ref{two}), we study the response of the Sakai-Sugimoto model to a constant electric background by turning it on 
inside the world volume of the probe $D8/\bar{D8}$ brane. The DBI action of the
 probe $D8/\bar{D8}$ brane is modified accordingly. Moreover, we systematically 
consider some time dependent fluctuation over the constant electric background and construct the associated effective classical
Lagrangian as a leading Maxwell term in the $\alpha'$ expansion of the modified DBI action. Our aim is to add an one-loop 
effective action to this classical Maxwell action and study the gauge field dynamics governed by the total action. 
It is important to mention that the quantum action we are interested in can be obtained by systematically covariantizing the one in 
four dimensional flat space time. Therefore to accommodate 
the quantum action we need to dimensionally reduce the classical action to a four dimensional subspace of the 
$D8/\bar{D8}$ world volume. 
The systematic recipe for adding the one loop effective Lagrangian capturing the physics of Schwinger pair production is 
described in section (\ref{three}). The Schwinger pair of our interest is the quark antiquark pair
associated with end points of the fundamental string joining either $D4-D8/\bar{D8}$ or $D8-\bar{D8}$ brane stacks (modulo the comments of the previous paragraph). 
Furthermore, in section (\ref{four}) by analyzing the quantum corrected effective Lagrangian, 
we demonstrate the holographic method to compute the conductivity in the strongly coupled 
boundary theory. For the conductivity computation we require the information of retarded Green's function. 
In this work we calculate the Green's function using numerical methods. 
Finally we plot the real part of the conductivity with the frequency of the fluctuating gauge field. 
We summarize our results and conclude with some remarks on our analysis in section (\ref{five}).

\section{Sakai-Sugimoto model} \label{one}
Sakai-Sugimoto model provides a very successful holographic method to construct the type 
IIA supergravity background, dual to the large $N_c$ QCD theory \cite{SakSug}. 
The IIA background includes a stack of $N_c$ color $D4$ branes wrapping a $S^1$ circle.
In addition to that, within probe approximation, $N_f$ number of flavor $D8/\overline{D8}$
branes at two different points on the circle are also considered, accounting no back reaction on the $D4$ geometry.
The four dimensional boundary theory 
of the $D4$ branes is realized as a QCD-like theory
with the $ N_c$ colors of gluons and the $N_f$ flavors of massless chiral fermions ($N_f \ll N_c$) with anti-periodic 
boundary condition. The left (right) handed fermions come from the 
spectrum of $D8(\overline{D8})$ flavor branes and they transform in the fundamental representation of both the color gauge group 
$U(N_c)$ and the flavor gauge group $U(N_f)_L\times U(N_f)_R$. The above realization of the Sakai-Sugimoto model 
is valid at a energy scale far below the Kaluza- Klein mass scale.

In more detail, we consider the near horizon limit of the 
$D4$ branes in the type IIA string theory. 
If we set the curvature of space time $R = {(\pi g_s N_c)}^{1/3} l_s = 1$ and impose the limit $N_c >>1$, 
the metric structure takes the following form,
\begin{eqnarray}
ds^2 = u^{3/2}(- dt^2 + dx^2 + dy^2 + dz^2 + f(u) dx_4^2) + u^{-3/2}(\frac{du^2}{f(u)}+u^2 d\Omega_4^2), \nonumber \\
f(u) = 1-(\frac{{u_{KK}}^3}{u^3}), ~~ e^{\Phi} = g_s u^{\frac{3}{4}}, ~~F_4 = 3 \pi l_s^3 N_c d\Omega_4,
\label{back}
\end{eqnarray}
where $u_{KK}^{-1/2} = \frac{3}{2} R_4$ signifies the  
supersymmetry breaking scale of the theory, $x_4$ is the circular direction with the periodicity, 
$x_4 \sim x_4 + 2\pi R_4$ and $\Omega_4$ is the volume element of $S^4$.
The $(u, x_4)$ subspace turns out to be topologically a cigar. The tip of this $(u, x_4)$ subspace is localized at $u =u_{KK}$. 
The $AdS/CFT$ correspondence maps this particular background (\ref{back}) to a 
confined phase of the boundary gauge theory at zero temperature with the 't Hooft coupling parameter,
\begin{eqnarray}
 \lambda_{Sakai-Sugimoto} = 4 \pi g_s N_c l_s.
 \label{thooft}
\end{eqnarray}
Possible embeddings of $D8$ and $\overline{D8}$ probe branes consistent with the topology of the background (\ref{back}) 
correspond to the various chiral phases in the boundary QCD like theory. For example, one can construct a $\mathcal{U}$ shaped embedding 
by smoothly joining the $D8$ and $\overline{D8}$ branes at 
a radial distance $u_0 \geq u_{KK}$. The $\mathcal{U}$ shaped embedding holographically conceives a very 
simple geometrical picture of the spontaneous breaking of 
the chiral gauge group $SU(N_f)_R \times SU(N_f)_L$  to the diagonal one $U(N_V)$ \cite{SakSug}. 
The form of $\mathcal{U}$ shaped embedding can be extracted by solving the equation of motion for $x_4$ obtained from the DBI action.
\begin{eqnarray}
 S_{D8}  =  -{\mathcal{T}}_8 \int d^9 X  e^{-\phi} Tr \sqrt{-\text{det}(\mathcal{G} +  2\pi l_s^2\mathcal{F})},
\end{eqnarray}
where ${\mathcal{T}}_8$ is the tension of $D8$ brane, $\mathcal{G} $ is the induced world volume of 
metric and $\mathcal{F}$ is the world volume gauge field.
The boundary condition for the solution is fixed by specifying that the asymptotic separation between the 
stack of $D8$ and $\overline{D8}$ branes at boundary is $L$. 
Using the background metric (\ref{back}), the DBI action of $D8$ and $\overline{D8}$ stacks reduces to,
\begin{eqnarray}
 S_{D8} = - \mathcal{W} \int du ~u^4 ~{\Big[ f(u) {x'_4}^2 + \frac{1}{u^3 f(u)}\Big]}^{\frac{1}{2}}.
\end{eqnarray}
Here we have introduced the normalization constant $\mathcal{W}$ as follows,
\begin{eqnarray}
 \mathcal{W} = 2 N_f V_{\mathcal{M}_4} \Omega_4 {\mathcal{T}}_8,
\end{eqnarray}
where $ V_{\mathcal{M}_4} $ is the volume of 4-dimensional space-time $(t,x,y,z)$.
Finally the $\mathcal{U}$ shaped profile is obtained by solving the equation of motion for $x_4(u)$ coordinate.
 \begin{equation}
x_4'(u) = {1\over u^{3/2} f(u)} \left[{u^8f(u)\over u_0^8 f(u_0)} - 1\right]^{-1/2} .
\label{prof1}
\end{equation}
 A finite temperature extension of the confining phase can be holographically realized 
by considering the Euclidean continuation of (\ref{back}). In the Euclidean signature, the time coordinate, $t^{E}$ 
develops a periodicity $t^{E} \sim t^{E}  + 1/T$, 
where $T$ is identified with the temperature of the gauge theory. Moreover, the gravity background dual to the deconfined phase at finite 
temperature is achieved with interchanging the role of $t$ and $x_4$ \cite{KS}. 
In Minkowski signature, the corresponding black hole background reads as, 
\begin{eqnarray}
ds^2 = u^{3/2}(- f(u) dt^2 + dx^2 + dy^2 + dz^2 +  dx_4^2) + u^{-3/2}(\frac{du^2}{f(u)}+u^2 d\Omega_4^2).
\label{backblack}
\end{eqnarray}
The background develops a horizon at $u = u_{T} = {(\frac{4 \pi T}{3})}^2$ and the blackening function is defined as, 
\begin{equation}
 f(u) = 1- \frac{u_T^3}{u^3}.
\end{equation}
The induced metric of $D8$ and $\overline{D8}$ world volume becomes,
\begin{equation}
ds^2_{D8}= u^{3/2}(-f(u) dt^2 + dx^2 + dy^2 + dz^2 ) + u^{-3/2}\Big(\frac{(1+u^3 f(u) x_4'^{2})}{f(u)}du^2 + u^2 d\Omega_4^2 \Big).
\label{ind1}
\end{equation}
In this blackhole background, the $(u, x_4)$ subspace is topologically a cylindrical 
and the possible embeddings of $D8/\overline{D8}$ brane
consistent with this topology turn out to the $\mathcal{U}$ shaped (broken chiral symmetry, $T < \frac{0.154}{L}$) and the parallel 
embeddings (restoration of the chiral symmetry, $T > \frac{0.154}{L}$) \cite{SakSug}. 
Again, both of them can be obtained from the DBI action with the boundary condition previously mentioned. 
\begin{eqnarray}
 S_{D8} =  -\mathcal{W} \int du ~u^4 ~{\Big[ f(u) {x'_4}^2 + \frac{1}{u^3}\Big]}^{\frac{1}{2}}.
\end{eqnarray}

The form of $\mathcal{U}$ shaped embedding is fixed by the following equation of motion of the
$x_4$ coordinate,
 \begin{equation}
x_4'(u) = {1\over u^{3/2} \sqrt{f(u)}} \left[{u^8f(u)\over u_0^8 f(u_0)} - 1\right]^{-1/2} .
\label{prof2}
\end{equation}

\begin{figure}[h]
 \centering
 \mbox{\subfigure[]{\includegraphics[width=3.5 cm]{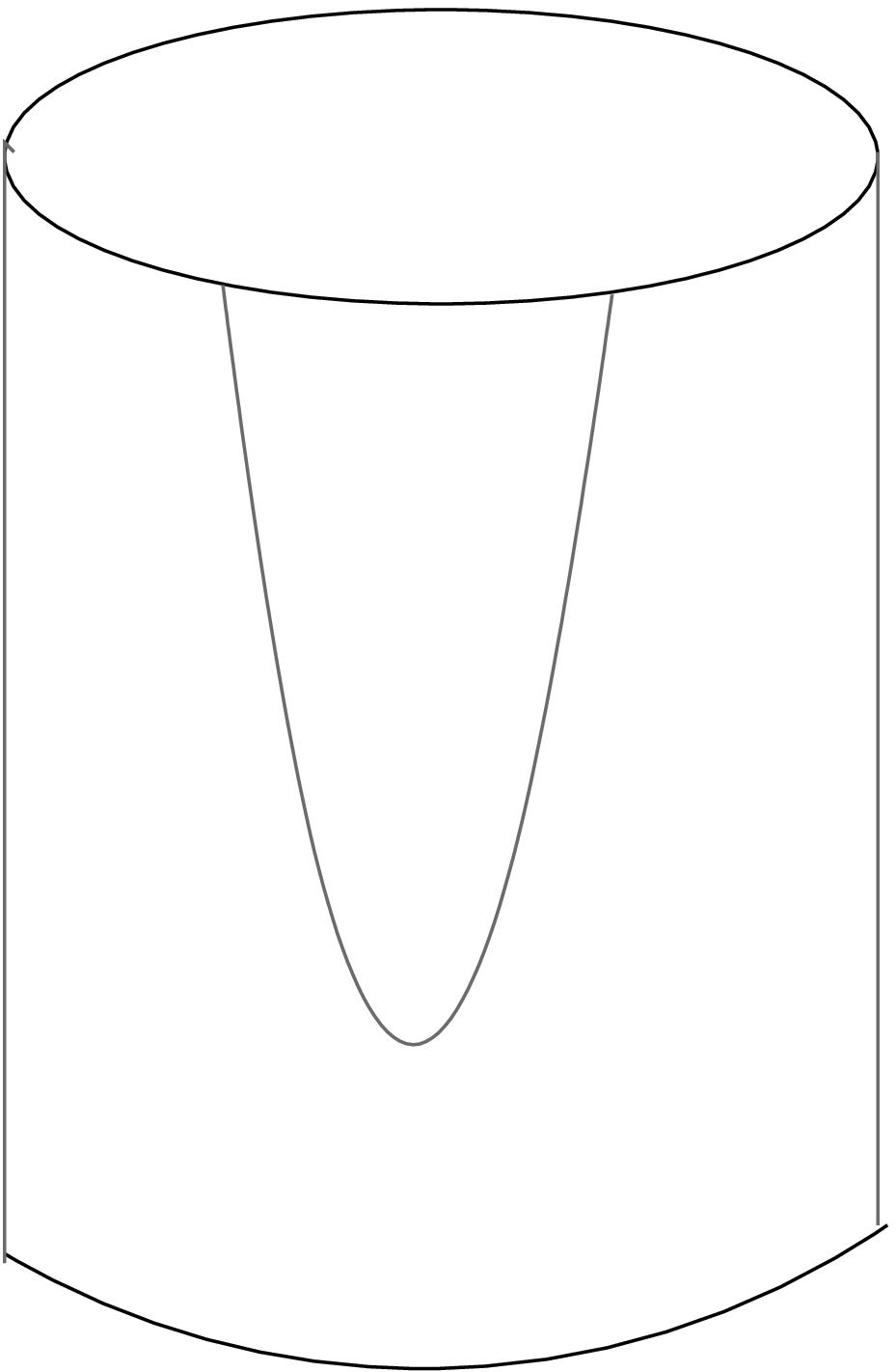}}
 \quad
 \subfigure[]{\includegraphics[width=3.5 cm]{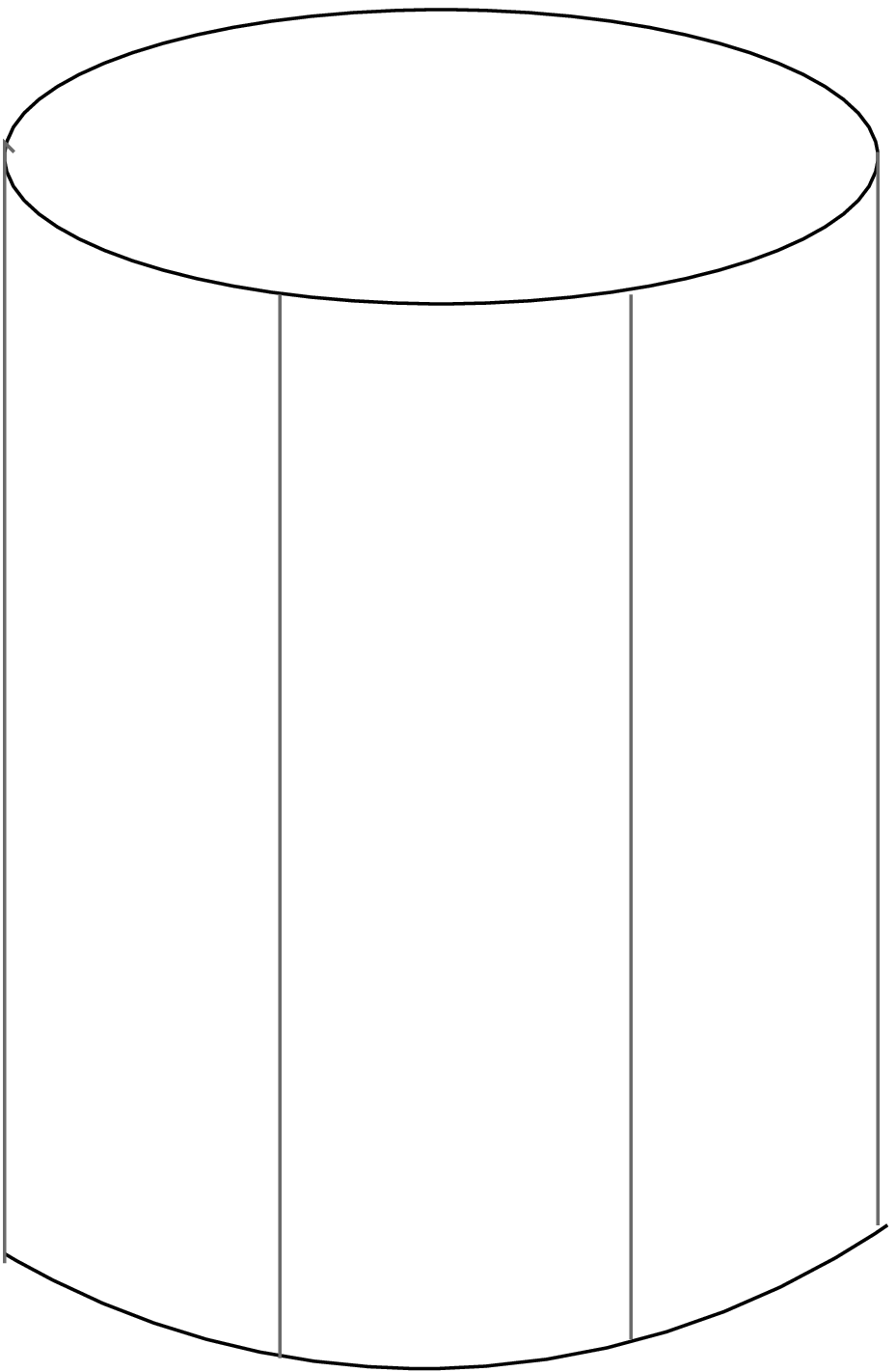} }}
 \caption{Here we show the embeddings of the $D8$ brane corresponding to appropriate boundary phases. 
Fig (a) corresponds to a deconfined boundary phase with broken chiral symmetry. Fig (b) depicts a deconfined boundary phase
with chiral symmetry restored.}
 \label{embed}
 \end{figure}
The allowed parallel $D8/\overline{D8}$ embedding satisfies,
 \begin{equation}
x_4'(u) = 0.
\label{prof12}
\end{equation}
The transition between the chiral symmetry broken phase to the 
chiral symmetry restored phase is first order. In the dual gravity theory, 
the curvature of $D8/\overline{D8}$ brane plays the role of order parameter. 
Following these holographic aspects of the 
Sakai-Sugimoto model, in the next section we elaborate upon the response of the deconfined phase with broken chiral symmetry to an 
external electric field.

\section{External electric field and the construction of effective Lagrangian}\label{two}
Recently in \cite{Bergmann}, authors have thoroughly studied the response Sakai-Sugimoto model to the external electric as well 
as magnetic field. Without the loss of generality we consider the electric field is directed along $z$ axis. 
The response to the external field is studied by turning on an Abelian component of the unbroken 
$U(N_V)$ gauge field in the world volume of probe $D8/\overline{D8}$ brane. 
\begin{eqnarray}
 \mathcal{A} = \frac{1}{N_f} tr \mathcal{A}_{N_V}.
\end{eqnarray}
In this analysis, assuming a current flowing along the direction of the electric field one can set 
a suitable choice of ansatz as,
\begin{equation}
 \mathcal{A}_z = E t + \Xi(u),
\end{equation}
where $E$ is the constant electric background and $\Xi(u)$ is the time-independent 
fluctuation encoding the holographic 
notion of boundary current \cite{Karch-metallic}. 
With the above choice of gauge field, the modified DBI action for the $D8/\bar{D8}$ brane embedded in the gravity 
background dual to the 
deconfined phase of finite temperature gauge theory, is given as 
\begin{eqnarray}
 S_{D8} = -\mathcal{W} \int du u^4 \sqrt{(f(u){x'_{4}}^2 + \frac{1}{u^3})(1-\frac{e^2}{f(u) u^3}) + \frac{f(u)\xi'^2}{u^3}},
\label{actiond8}
\end{eqnarray}
where we have introduced the dimensionless quantities like $e = 2 \pi l_s^2 E$ and $\xi = 2 \pi l_s^2 \Xi$. 

It is important to note that the DBI action does not
explicitly depend on the gauge field fluctuations. Therefore the equation of motion for $\xi$ evokes a constant of motion $\mathcal{J}$.
\begin{eqnarray}
 \frac{u f(u) \xi'}{\sqrt{(f(u){x'_{4}}^2 + \frac{1}{u^3})(1-\frac{e^2}{f(u) u^3})}} = \mathcal{J}.
\label{eomd8}
\end{eqnarray}
Using (\ref{eomd8}) and substituting $\xi'$ in terms of the constant of motion $\mathcal{J}$ in (\ref{actiond8}) we get,
\begin{eqnarray}
  S_{D8} = -\mathcal{W} \int du u^4 \sqrt{\Big(f(u){x'_{4}}^2 + \frac{1}{u^3}\Big)\Big(f(u)-\frac{e^2}{u^3}\Big) 
  {\Big(f(u)-\frac{{\mathcal{J}}^2}{u^5}\Big)}^{-1}}.
\label{actiond81}
\end{eqnarray}
Moreover, by demanding the reality of the DBI action (\ref{actiond81}) the response coefficient like conductivity
is holographically computed in the boundary theory.
By integrating (\ref{eomd8}) with respect to $u$ and then taking the boundary limit, the result 
gives the leading asymptotic behavior of the gauge field,
\begin{eqnarray}
2 \pi l_s^2 \mathcal{A}_z = e t  -\frac{2}{3} \frac{\mathcal{J}}{u^{\frac{3}{2}}},
\end{eqnarray}
where $\mathcal{J}$ can be physically interpreted as the conserved current associated with $\mathcal{A}_z$.
To obtain the $\mathcal{U}$ shaped embedding of $D8/\bar{D8}$ brane we solve 
the equation motion for $x'_{4}$ derived from the DBI action (\ref{actiond8}).
 \begin{equation}
x_4'(u) = {1\over u^{3/2} \sqrt{f(u)}} \left[\frac{u^8f(u)(f(u)-\frac{e^2}{u^3}){(f(u)-\frac{\mathcal{J}^2}{u^5})}^{-1}}
{ u_0^8 f(u_0) 
(f(u_0)-
\frac{e^2}{u_0^3}){(f(u_0)-\frac{\mathcal{J}^2}{u_0^5})}^{-1}} 
- 1\right]^{-1/2}.
\label{prof3}
\end{equation}
It is straightforward to get an on-shell DBI action $S^{\mathcal{U}}_{\mathcal{J}\ne 0}$ by substituting the above solution into 
(\ref{actiond8}). 
However, in \cite{Bergmann}, it has been shown that the physically 
favored configuration allows no current as the on-shell DBI action always satisfies
$S^{\mathcal{U}}_{\mathcal{J}\ne 0} > S^{\mathcal{U}}_{\mathcal{J} = 0}$, 
where $S^{\mathcal{U}}_{\mathcal{J} = 0}$ can be constructed by setting $\mathcal{J} = 0$ in (\ref{prof3}) and then substituting it 
back to (\ref{actiond8}). 
Therefore, in the boundary theory the deconfined phase with broken chiral symmetry
turns out to be insulating.

In our work, we accomplish a holographic re-analysis of the 
aforementioned boundary phase but with an explicit time dependence in the
gauge field fluctuation inside the $D8/\bar{D8}$ world volume. 
For further computation, we consider an approximation that the generalization of the time independent gauge field 
fluctuation into the time-dependent one will not modify the $\mathcal{U}$ shaped classical embedding for 
$\mathcal{J} = 0$.
The field ansatz in the Minkowski signature we are considering reads as 
\begin{equation}
\mathcal{A}_z =  E t + \tilde{\Xi}(t, u).
\label{ans}
\end{equation}
Using this time-dependent gauge field ansatz (\ref{ans}), 
it is straightforward to write down the DBI action for the $D8/\bar{D8}$ brane. We will purposefully
factorize the action for our future convenience. In particular, we shall keep track of the electric field by redefining
 the term under first square root as the modified world volume metric and then derive the fluctuation Lagrangian.
\begin{eqnarray}
S_{D8}  &=& -\frac{2 N_f \mathcal{T}_{D8}}{g_s} \int d^9x ~u^{-3/4}\sqrt{-Det(\mathcal{G}+(\mathcal{E} +\tilde{\mathcal{X}}(t, u)))}  \nonumber \\
       &=&  -\frac{2 N_f \mathcal{T}_{D8}}{g_s} \int d^9x ~u^{-3/4} \sqrt{-Det(\mathcal{G} +  \mathcal{E} )}  \nonumber \\
&& ~~~~~~~~~~~~~~~~~~~~~~\times \sqrt{Det(1 +  \tilde{\mathcal{X}}(t, u)(\mathcal{G} +  \mathcal{E})^{-1})}.
\label{newaction}
\end{eqnarray}
In our notation, $\mathcal{G}$ is again the world volume metric tensor for the $D8/\bar{D8}$ brane. $\mathcal{E}$ and $\mathcal{X}(t, u)$ 
are the world volume field strength tensors for the constant background and the time-dependent fluctuation respectively. 
The non-zero components of $\mathcal{E}$
are 
\begin{eqnarray}
 \mathcal{E}_{tz} = -\mathcal{E}_{zt} = e,
\end{eqnarray}
whereas, the same for $\mathcal{X}(t, u)$ are 
\begin{eqnarray}
 \mathcal{X}_{tz} = - \mathcal{X}_{zt}= \dot{\tilde{\xi}}(t,u),~~ \mathcal{X}_{uz} = - \mathcal{X}_{zu}= \tilde{\xi}'(t,u).
\end{eqnarray}

Similar to the time independent case, here we also introduce the dimensionless quantities, 
$e = 2 \pi l_s^2 E$ and $\tilde{\xi} = 2 \pi l_s^2 \tilde{\Xi}$.
For small values of gauge field fluctuation, we expand the DBI action using the following matrix identity,
\begin{equation}
 \sqrt{Det(1+M)} = 1 + \frac{1}{2}Tr M + \frac{1}{8} ({Tr M}^2) - \frac{1}{4} Tr{(M)}^2,
\end{equation}
where $M$ stands for $\tilde{\mathcal{X}}(t, u)(\mathcal{G} + \mathcal{E})^{-1}$. 
In this expansion, it is sufficient for our purpose to keep the terms up to the quadratic fluctuations.

Neglecting those terms which are either constant or linear in fluctuation, thus not contributing to the equation of motion
 and putting everything else together up to quadratic order we get the classical action for fluctuation,
\begin{eqnarray}
S_{\text{cl}} &=& -\frac{2 N_f \mathcal{T}_{D8}}{g_s}\int d^9 x~ u^{-3/4} \sqrt{f(u)u^{3/2} 
(1-\frac{e^2}{u^3 f(u)}) {(u^{3/2})}^3 ~\frac{u^{3/2}}{f(u)}
 (\frac{1}{u^3}+f{x_4'}^2)
\mathcal{G}_{\Omega_4\Omega_4}}\nonumber \\
&&  \times  \Big(- \frac{{\dot{\tilde{\xi}}}^2}{2f(u)u^3{(1-\frac{e^2}{f(u)u^3})}^2} + \frac{f(u){\tilde{\xi'}}^2}{2u^3(1-\frac{e^2}{f(u)u^3})
(\frac{1}{u^3} +f(u){x_4'}^2)}\Big) .\nonumber \\
\label{eqn1}
\end{eqnarray}
Our aim is to write $S_{cl}$ obtained by expanding DBI action up to quadratic order of fluctuation 
in a canonical four dimensional Maxwell form. The motivation for imposing the restriction on space time dimension comes from the 
fact that at the end we aim to study the effect of Schwinger pair production on electrical conductivity in a strong coupling phase 
by analyzing the dynamics of gauge field fluctuation governed by a quantum corrected action and the quantum contribution for the 
action we are interested in is originally constructed in \cite{Urrutia} for four dimensional flat space time. 
Therefore to successfully accommodate the quantum correction we need the above mentioned dimensional reduction.
To do so we proceed in the following way. First, by observing the factors under the square root in equation (\ref{eqn1}) we 
make an ansatz for a modified induced metric $\mathcal{\tilde{G}}$. 
This modified metric absorbs the extra factor coming due to the presence of constant electric background $E$ 
and reduces to the original one, $\mathcal{G}$ 
in the $E \rightarrow 0$ limit. 
Then we execute a dimensional reduction 
on the DBI action realized over nine dimensional world volume 
coordinates ($t,x,y,z,u,\theta_1,\theta_2,\theta_3,\theta_4$) and write it as an effective action in four dimensional space time 
($t,y,z,u$), where $\theta_i$'s are the 
angular variables in $\Omega_4$. Finally we redefine the coupling parameter in the four dimensional effective action 
in such a way the final expression takes the desired canonical form.
The ansatz for the modified modified induced metric $\mathcal{\tilde{G}}$ of our interest is set as,
\begin{eqnarray}
 \mathcal{\tilde{G}}_{tt} &=& (1-\frac{e^2}{f(u)u^3})\mathcal{G}_{tt}, \nonumber \\
\mathcal{\tilde{G}}_{ij} &=& \mathcal{G}_{ij} ~~~~~~~\forall i,j.
\label{modmet}
\end{eqnarray}
With this modified metric (\ref{modmet}), we re-write the $S_{cl}$ in (\ref{eqn1}) in the following way,
\begin{eqnarray}
 S_{\text{DBI}} &=&  -\frac{2 N_f \mathcal{T}_{D8}}{g_s}\int d^9 x~ u^{-3/4} \sqrt{-Det\mathcal{\tilde{G}}} ( \frac{1}{1-\frac{e^2}{f(u)u^3}})~~\frac{1}{2}
\{\mathcal{\tilde{G}}^{tt}\mathcal{\tilde{G}}^{zz} { \mathcal{X}_{tz}}^2 + \mathcal{\tilde{G}}^{uu}\mathcal{\tilde{G}}^{zz} 
{ \mathcal{X}_{uz}}^2  \}. \nonumber \\
\label{eqn2}
\end{eqnarray}
Moreover, we perform the dimensional reduction from nine to four dimensions. 
It is important to note that in 
doing so we make another assumption that one of the spatial boundary coordinates is wrapped in a 
circle and introduces another energy scale $< u_{KK}$.
For all subsequent part of this paper we restrict our analysis far below this newly introduced energy scale. 
Redefining the coupling constant in a 
suitable way we finally land up with a four dimensional effective action in a canonical Maxwell form.
\begin{eqnarray}
 S_{\text{Classical}} &=& -\int d^4x \frac{1}{g_s^{eff}} \sqrt{-Det{\mathcal{\tilde{G}}}_4} ~~\frac{1}{4} F_4^2, 
\label{classical1}
\end{eqnarray}
where, $g_s^{eff} = \frac{g_s (1-\frac{e^2}{f(u)u^3})}{2N_f \mathcal{T}_{D8}V_{\Omega_4} V_x 4 \pi^2l_s^4 u} = 
\hat{g}_s^{eff}\frac{1-\frac{e^2}{f(u)u^3}}{u}$.
${\mathcal{\tilde{G}}}_4$ is the four dimensional effective metric having non-zero diagonal element,
\begin{eqnarray}
 {({\mathcal{\tilde{G}}}_4)}_{t t} = {\mathcal{\tilde{G}}}_{t t}, ~~  {({\mathcal{\tilde{G}}}_4)}_{u u} = {\mathcal{\tilde{G}}}_{u u }, ~~
   {({\mathcal{\tilde{G}}}_4)}_{yy } = {\mathcal{\tilde{G}}}_{y y},  ~~    {({\mathcal{\tilde{G}}}_4)}_{z z} = {\mathcal{\tilde{G}}}_{zz}. 
\label{modmet1}
\end{eqnarray}

Since $u$ has been chosen to have zero length dimensions, the running Yang-Mills coupling constant consistently turns out to be dimensionless. 
The non-zero components of the four dimensional field strength tensor $F_4$ are the following,
\begin{eqnarray}
 {F_4}_{t z} = \dot{\tilde{\Xi}}(t, u), ~~~~  {F_4}_{u z} = \tilde{\Xi}'(t, u).
\label{4dimfield}
\end{eqnarray}
Up to this stage, we construct an effective Maxwell action for gauge field fluctuation around a constant electric background 
in four dimensional subspace of  $D8/\bar{D8}$ brane world volume characterized by the modified metric 
${\mathcal{\tilde{G}}}_4$. In the next section, we discuss how to add a four dimensional effective action arising from the one-loop 
correction.

\section{Adding one-loop quantum correction} \label{three}
In this section we discuss how to add a consistent one loop effective action to the tree level canonical Maxwell action, 
(\ref{classical1}) in four dimensional space time defined by $\tilde{\mathcal{G}}_4$. 
Following the methodology of QED, we point out that the structure of quantum 
action, $S^{Q}$ solely depends on the loop contribution arising from the vacuum polarization tensor $M_{\mu\nu}$, 
generically prescribed as a two point correlator of the propagating 
gauge field. 
In this paper we consider the modification of gauge field propagation in vacuum up to one loop order. The modification is due to 
relevant polarization tensor $M_{\mu\nu}$ encoding the process of virtual quark-antiquark pair creation. 
In the presence of an external homogeneous electric field $E$, the polarization tensor develops an explicit 
dependence on the respective electric
field and further restricts the vacuum dynamics. Motivated by Schwinger's seminal work \cite{Schwinger:1951nm}, 
there has been already an extensive perturbative as well as non-perturbative study to compute the 
vacuum polarization tensor for the pair creation process in terms 
of double parameter integral, including the presence of a background electromagnetic field in four dimensional Minkowski space
\cite{Urrutia, Tsai:1974fa, Tsai:1975iz, Baier:2007dw, Baier:2009it, Dittrich:1985yb, Karbstein:2013ufa}. 

In this kind of computations the externally set vector parameters are the field vectors and the transferred four momentum vector
$k_{\mu}$. For the simplification of the computation it is always convenient to decompose the back ground space time
along the parallel and the perpendicular directions to the external field. Without the loss of generality we consider the 
electric field is directed along $z$ axis.
The four dimensional Minkowski space time coordinates $(t, y, u ,z)$, can be grouped as $(t, z) \Rightarrow ||$, and 
$(u, y) \Rightarrow \perp$. Subsequently, the metric can be decomposed in the following way,
\begin{eqnarray}
 \eta^{\mu\nu} =  \eta^{\mu\nu}_{||} + \eta^{\mu\nu}_{\perp}, 
\end{eqnarray}
where $||/\perp$ stands for $(t, z)/(y, u)$ respectively. Furthermore, the transferred four momentum can be decomposed as,
\begin{eqnarray}
 k^{\mu} =  k^{\mu}_{||} + k^{\mu}_{\perp}, 
\end{eqnarray}
With above consideration, following \cite{Urrutia}, 
the momentum space representation of the polarization tensor in the presence of a homogeneous external field can be expressed up to 
one loop level as,
\begin{eqnarray}
M^{\mu \nu} &=& \frac{\mathcal{C}}{2}\int_0^{\infty} \frac{ds}{s}\int_{-1}^{1}\frac{dv}{2} e^{-is\phi_0} \frac{ w^{'}}{\sinh w^{'}} \Big[(\eta^{\mu \nu}k^2 - k^{\mu}k^{\nu}) N_0 
+ (\eta_{||}^{\mu \nu}k_{||}^2 -  k_{||}^{\mu}k_{||}^{\nu}) N_1 +
  \nonumber \\
&& ~~~~~~~~~~~~~~~~~~~~~~~~~~~~~~~~~~~~~~~~~~~~~~~~~~~~~~~~ + \text{contact terms} \Big],
\label{vacpol}
\end{eqnarray}
where $\mathcal{C}$ is an overall normalization constant and the other parameters $\phi_0$, $N_0 $ and $N_1$ are defined as,
\begin{eqnarray}
  N_0 &= &\cosh w^{'}v -v \sinh w^{'}v  \coth w^{'},
\label{para1} \\
  N_1& = &2  \frac{\cosh w^{'} - \cosh w^{'}v }{\sinh^2 w^{'}} - N_0,
\label{para2}  \\ 
 s \phi_0 &= &   sm^2 + \frac{k_{||}^2}{2} \frac{\cosh w^{'} - \cosh w^{'}v}{z^{'}\sinh w^{'}}
+\frac{1}{4} s (1-v^2)k_{\perp}^2 ,
\label{para3}\\ 
w^{'}& = & q s E,
\label{para4}
 \end{eqnarray} 
where $m$ and $q$ are the mass and charge of the flavor quark. The contact term  in (\ref{vacpol}) can be fixed by demanding the following two conditions,
\begin{eqnarray}
 k^2 = 0,  ~~~~~~~  
\lim_{\substack{ k^2 \to 0 \\ E \to  0}} M_{\mu \nu} = 0.
\end{eqnarray}
The form of contact term turns out to be,
\begin{eqnarray}
 \text{contact term} = - (1-v^2)(\eta^{\mu \nu}k^2 - k^{\mu}k^{\nu}).
\label{contact}
\end{eqnarray}
The constraints , $k^2 = 0$ on the regime of transferred four momentum signifies that the derivation of the $M_{\mu\nu}$ is fitted
to on-the-light-cone dynamics. 
There are other possible forms of contact term which can be derived on-the-light-cone. However we have observed that the symmetry of frequency
dependent conductivity due to the pair creation is heavily dependent on the form of contact term as it contributes to 
the quantum action. 
Moreover, the holographic realization of the linear response theory suggests that the real part of the frequency dependent conductivity should be 
symmetric with respect to 
the associated frequency. By accounting these two facts we fix the contact term contributing to the polarization tensor to the above form 
and the one-loop action in momentum representation becomes,
\begin{equation}
{ \mathcal{S}}^Q = - \int \frac{d^4k}{{(2\pi)}^{4}} \sqrt{- Det\eta}  A^{\mu}(k)M_{\mu\nu} (k)A^{\nu}(-k),
\label{qlag}
\end{equation}
where $M_{\mu \nu}$ is specified by the equations (\ref{vacpol}), (\ref{para1}), (\ref{para2}), (\ref{para3}), (\ref{para4}) and 
(\ref{contact}). Translating from the momentum representation to the space time 
representation the quantum action simplifies into,
\begin{eqnarray}
{\mathcal{S}}^Q &=& - \frac{\mathcal{C}}{4}  \int d^4x \sqrt{-Det \eta} \Big[Tr ({(F_4)}_{||}.\eta_{||}.{(F_4)}_{||}.\eta_{||})( I_1) 
\nonumber \\
&& ~~~~~~~~~~~~~~~~~~~~~~~~~~~~~~~~ + 2Tr ({(F_4)}_{mixed}.\eta_{||}{(F_4)}_{mixed}\eta_{\perp})(I_2) \Big],
\nonumber \\
\end{eqnarray}
where $F_4$ is the same second rank field strength tensor of propagating gauge field defined in (\ref{4dimfield}). 
More specifically, ${(F_4)}_{||}$ / ${(F_4)}_{\perp}$ 
allows only those space time indices which signify the
parallel/perpendicular direction to the external homogeneous electric field. On the other hand, ${(F_4)}_{mixed}$ signifies that one of 
its space time 
indices belongs to $(t, z) \Rightarrow ||$ and the other one takes value in $(u, y) \Rightarrow \perp$. Furthermore, $
I_1$ and $I_1$ carry the information of the double integrals in the action.
\begin{eqnarray}
 \nonumber I_1 = \int \frac{ds}{s}\int \frac{dv}{2} e^{-i s \phi_0}  \frac{ w^{'}}{\sinh w^{'}}(N_0+N_1-(1-v^2)), \nonumber \\
I_2 = \int \frac{ds}{s}\int \frac{dv}{2} e^{-i s \phi_0}  \frac{ w^{'}}{\sinh w^{'}} (N_0-(1-v^2)). 
\label{integralcouple}
\end{eqnarray}

However, we need to construct a one-loop quantum corrected action in the four dimensional curved space time denoted by 
$\tilde{\mathcal{G}}_4$.  Covariantization is to be done by using Riemann normal coordinates (RNC)\cite{Pet,Eis}. 
If we denote by $y^\mu$ the RNC, then one interprets $k_\mu$ as $-i {\partial \over \partial y^\mu}$. 
Various powers of derivatives can be re-expressed in terms of covariant derivatives and  the curvature tensors. 
If the radius of curvature of the space is large enough, then to a first approximation we can neglect the curvature terms. 
This is what will be done in this paper.  In the large $N_c$ limit, we can consider only the leading term for simplification of our analysis.
\begin{eqnarray}
{ \mathcal{S}}^Q &=& -\frac{\mathcal{C}}{4}  \int d^4x \sqrt{-Det \tilde{\mathcal{G}}_4}
 \Big[Tr {(F_4)}_{||}.{(\tilde{\mathcal{G}}_4)}_{||}.{(F_4)}_{||}.{(\tilde{\mathcal{G}}_4)}_{||})( I_1) + \nonumber \\
&& ~~~~~~~~~~~~~~~~~~~~~~~~~~~~~~~~
2Tr ({(F_4)}_{mixed}.{(\tilde{\mathcal{G}}_4)}_{||}{(F_4)}_{mixed}{(\tilde{\mathcal{G}}_4)}_{\perp})(I_2) \Big], \nonumber \\
\label{covaction}
\end{eqnarray}
where we consider the contravariant components of the 
metric tensor $\tilde{\mathcal{G}}_4$ and the covariant component of the field tensor $F_4$. 
It is extremely difficult to evaluate the integrals $I_1$ and $I_2$ with full generality. One of the ways out is to constrain the 
full 
parameter space by imposing weak field approximation,
\begin{eqnarray}
 \frac{q E}{m^2} <<1.
\end{eqnarray}
Within this weak field approximation if we expand the r.h.s of (\ref{para3}) in the powers of $E$, the 
leading order contribution becomes: (We are interested in the frequency dependence of the conductivity. 
Currents are assumed to be time varying but uniform in space. So we set $\vec k=0$ and $ k_0 = \omega \neq 0$.),
\begin{eqnarray}
s\phi_0 = sm^2 +\frac{1}{48}q^2\omega^2 |{\mathcal{\tilde{G}}}_{00}| s^3 {(1-v^2)}^2E^2.
\label{phiapprox}
\end{eqnarray}
We can define a relevant dimensionless parameter $\lambda$ such that it takes both small and large numerical 
values depending on the 
magnitude of frequency and the intensity of the field.
\begin{eqnarray}
 \lambda = \frac{3}{2} \frac{q E}{m^2}\frac{w}{m}.
\end{eqnarray}
 Similarly, under same approximation
\begin{eqnarray}
 (N_0 + N_1-(1-v^2)) \frac{ w^{'}}{\sinh w^{'}} =    - \frac{1}{12} E^2 q^2 s^2 {(5-6v^2+v^4)},  \nonumber \\
(N_0-(1-v^2)) \frac{ w^{'}}{\sinh w^{'}} =   - \frac{1}{6} E^2 q^2 s^2 {(1-v^2)}^2.
\label{const1}
\end{eqnarray}
It is important to note that, weak field approximation does not make contradiction with large and small values of $\lambda$. In both 
cases the leading contribution for the integral mentioned in (\ref{vacpol}) comes from $s<<1$ \cite{Tsai:1974fa}.

Plugging the results of (\ref{phiapprox}) and (\ref{const1}) in (\ref{integralcouple}) we get,
\begin{eqnarray}
 \nonumber I_1 &=&  - \frac{1}{12} E^2 q^2 \int \frac{dv}{2} {(5-6v^2+v^4)}  \int s ds  e^{-i s \phi_0(\omega)},
\label{i1} \\ 
I_2 &=&  - \frac{1}{6} E^2 q^2 \int \frac{dv}{2} {(1-v^2)}^2  \int s ds  e^{-i s \phi_0(\omega)}.
\label{i2},
\end{eqnarray}
Now we evaluate the integrals $I_1$ and $I_1$ within the weak field approximation. To do so, first we 
work out the $s$ integrals by applying the standard integral representation of the known special function. 
We introduce the variable $r$ and the parameter $\rho$ for simplification of the computation,
\begin{eqnarray}
 r = \frac{1-v^2}{4}\frac{w}{m}q E \sqrt{|{\mathcal{\tilde{G}}}_{00}|} ~~s,  \nonumber \\
\rho = \frac{4}{\lambda}\frac{1}{1-v^2}\frac{1}{\sqrt{{|\mathcal{\tilde{G}}}_{00}|}}.
\label{def}
\end{eqnarray}
With the above redefinition, the relation between the $s$ and $r$ can be written in a compact form,
\begin{eqnarray}
 s \phi_0 &=& \frac{3}{2}\rho (r + \frac{r^3}{3}) , ~~~~~~~s = {(\frac{3 \rho}{2 m^2})} r.
 \end{eqnarray}
Correspondingly the real and the imaginary part of the $s$ integral are successfully translated into respective components 
of $r$ integral.
\begin{eqnarray}
 \int_0^{\infty} s ds  \cos{(s \phi_0)} &=& 
\frac{1}{m^4} {(\frac{3 \rho}{2 })}^2 \int_0^{\infty}  r dr \cos(\frac{3\rho}{2}(r+\frac{r^3}{3})) 
\label{intres1},\\
\int_0^{\infty} s ds  \sin{(s \phi_0)} &=& 
\frac{1}{m^4} {(\frac{3 \rho}{2 })}^2 \int_0^{\infty}  r dr \sin(\frac{3\rho}{2}(r+\frac{r^3}{3})).
\label{intres2}
 \end{eqnarray}
Again a suitable change of variable $r = {(\frac{3 \rho}{2})}^{-1/3} p$, recasts the r.h.s of the equation 
(\ref{intres1}) in the desired form. 

\begin{eqnarray}
\int_0^{\infty}  r dr \cos\Big(\frac{3\rho}{2}(r+\frac{r^3}{3})\Big) &=& {(\frac{3 \rho}{2 })}^{-2/3}
\int_0^{\infty}  p dp \cos\Big({(\frac{3 \rho}{2 })}^{2/3}p +\frac{p^3}{3}\Big),
\nonumber \\
&=& \pi {(\frac{3 \rho}{2 })}^{-2/3} Gi'[{(\frac{3 \rho}{2 })}^{2/3}],
\end{eqnarray}
where $Gi[x]$ is the inhomogeneous Airy function/Scorer function satisfying the 
following differential equation and the boundary conditions \cite{Gil}.
\begin{eqnarray}
 Gi''[x] -xGi[x] &=& - \frac{1}{\pi}, \nonumber \\
Gi[0] &=& \frac{1}{3} Bi[0] = \frac{1}{\sqrt{3}} Ai[0], \nonumber \\
Gi'[0] &=& \frac{1}{3} Bi'[0] = -\frac{1}{\sqrt{3}} Ai'[0], \nonumber \\
\end{eqnarray}
where $Ai[x]$ and $Bi[x]$ are the homogeneous Airy function of the first kind and the second kind respectively.
Integral representation of $Gi[x]$ is given as ,
\begin{eqnarray}
 Gi[x] = \frac{1}{\pi} \int_0^{\infty}  dz \sin[x z +\frac{z^3}{3}].
\end{eqnarray}

Moreover, using integral representation of the modified Bessel function, 
the r.h.s of the equation (\ref{intres2}) can be expressed as,
\begin{eqnarray}
\int_0^{\infty}  r dr \sin[\frac{3\rho}{2}(r+\frac{r^3}{3})] = \frac{ K_{2/3}[\rho] }{\sqrt{3}},
\end{eqnarray}
where $K_{2/3}[\rho]$ is the generalized Bessel function with $2/3$ weight \cite{abram}.
Now combining all the results we can finally write down the result of $s$ integral calculation,
 \begin{eqnarray}
\int_0^{\infty} s ds  e^{-i s\phi_0} = 
 \frac{\pi}{m^4} {(\frac{3 \rho}{2 })}^{4/3}Gi'[{(\frac{3 \rho}{2 })}^{2/3}] - 
i {(\frac{3 \rho}{2 m^2 })}^{2} \frac{ K_{2/3}[\rho] }{\sqrt{3}}.\nonumber \\
\end{eqnarray}

For our present analysis we are interested in the asymptotic region of the associated parameter $\rho $. 
If we consider $\rho$ is very large ($\rho >>1$), the $s$ integral simplifies to,
\begin{eqnarray}
 \int_0^{\infty} s ds  e^{-i s\phi_0} = -\frac{1}{m^4} - i {(\frac{3 \rho}{2 m^2})}^2 \sqrt{\frac{\pi}{6 \rho}} e^{-\rho}.
\label{large}
\end{eqnarray}
Here we have used the following two asymptotic expansion \cite{abram},
\begin{eqnarray}
Gi'[x] \approx  - \frac{1}{\pi x^2}, \nonumber \\
K_{2/3}[x] \approx  \sqrt{\frac{\pi}{2 x}} e^{-x}.  \nonumber \\
\label{id5}
\end{eqnarray}
On the other hand, when $\rho$ takes small values ($\rho << 1$) the $s$ integral modifies as,
\begin{eqnarray}
 \int_0^{\infty} s ds  e^{-i s\phi_0} = \frac{\rho^{4/3}}{m^4} c_1,  ~~~~~
\Big(c_1 =  \Gamma[\frac{2}{3}] \Big(\frac{3}{2^{7/3}} -i \frac{3^{3/2}}{2^{7/3}} \Big) \Big).
\label{id6}
\end{eqnarray}
In deriving equation (\ref{id6}), we have used the following approximations \cite{abram},
\begin{eqnarray}
Gi'[x] \approx \frac{ 3^{-1/3} }{2 \pi}\Gamma[\frac{2}{3}], \nonumber \\
K_{2/3}[x] \approx 2^{-1/3}\Gamma[\frac{2}{3}] {x}^{-2/3}.  
\label{id7}
\end{eqnarray}
Having the $s$ integral done, we aim to work out the $v$ integral within the restricted range of parameters.
For small values of $\lambda$ we combine the results mentioned in (\ref{large}) and (\ref{i1}),
\begin{eqnarray}
 I_1 &=& \frac{4 q^2 E^2}{15m^4} + \nonumber \\
&& \frac{3 i q^2 E^2}{2\sqrt{6} m^4 |{(\tilde{\mathcal{G}}_4)}_{00}|} 
\sqrt{\frac{\pi \sqrt{|{(\tilde{\mathcal{G}}_4)}_{00}|}}{\lambda^3}}\int_0^1 dv \sqrt{1-v^2}~
\Big(\frac{5-6v^2+v^4}{{(1-v^2)}^2}\Big)~
e^{\frac{-4}{\lambda \sqrt{|{(\tilde{\mathcal{G}}_4)}_{00}|} (1-v^2) }}. \nonumber \\
\end{eqnarray}
Implementing the change of variable $x = \frac{1}{1-v^2}$ and applying the useful identity,
\begin{eqnarray}
 \int_{1}^{\infty} \frac{1}{\sqrt{x-1}}\frac{dx}{x^2}x^{\sigma}e^{-\delta x} \approx \sqrt{\frac{\pi}{\delta}}e^{-\delta}
\Big[1+ \mathcal{O}(\frac{1}{\delta}) \Big],
\end{eqnarray}
we get the final expression of $I_1$ for small $\lambda$,
\begin{eqnarray}
 I_1 &=& \Big(\frac{4 \omega_0^2}{15 m^2}\Big) +
 i\Big(  \frac{5 \pi} 
{2\sqrt{6} \sqrt{{\tilde{\mathcal{G}}}_{00}}} 
(\frac{\omega}{\omega_0}) \Big)e^{-\frac{4}{\lambda \sqrt{{\tilde{\mathcal{G}}}_{00}}}}~~~~(\omega_0 = \frac{qE}{m}). 
\end{eqnarray}
Similarly the $I_2$ integral turns out to be,
\begin{eqnarray}
 I_2 &=& \Big(\frac{4 \omega_0^2}{45 m^2}\Big) +
 i\Big(  \frac{\pi} 
{\sqrt{6} \sqrt{{\tilde{\mathcal{G}}}_{00}}} 
(\frac{\omega}{\omega_0})\Big)e^{-\frac{4}{\lambda \sqrt{{\tilde{\mathcal{G}}}_{00}}}}.
\end{eqnarray}
For large $\lambda$, same type of computation yields,
\begin{eqnarray}
 I_1 &=& -\Big[\Big( \frac{2^{\frac{5}{3}} \sqrt{\pi} \Gamma[\frac{5}{3}]\Gamma[\frac{2}{3}]}{3^{\frac{4}{3}} \Gamma[\frac{13}{6}]} {(\frac{m \omega_0}{{\tilde{\mathcal{G}}}_{00} \omega^2})}^{\frac{2}{3}}\Big) 
- i  \Big( \frac{2^{\frac{5}{3}} \sqrt{\pi} \Gamma[\frac{5}{3}]\Gamma[\frac{2}{3}]}{3^{\frac{5}{6}} \Gamma[\frac{13}{6}]} 
{(\frac{m \omega_0}{{\tilde{\mathcal{G}}}_{00} \omega^2})}^{\frac{2}{3}} \Big) \Big], \nonumber \\
I_2 &=& -\Big[\Big( \frac{2^{\frac{5}{3}} \sqrt{\pi} \Gamma[\frac{2}{3}]^2}{7 \times 3^{\frac{4}{3}} \Gamma[\frac{7}{6}]}{(\frac{m \omega_0}{{\tilde{\mathcal{G}}}_{00} \omega^2})}^{\frac{2}{3}}\Big) 
- i  \Big( \frac{2^{\frac{5}{3}} \sqrt{\pi} \Gamma[\frac{2}{3}]^2	}{7 \times 3^{\frac{5}{6}} 
\Gamma[\frac{7}{6}]} {(\frac{m \omega_0}{{\tilde{\mathcal{G}}}_{00} \omega^2})}^{\frac{2}{3}}
\Big)\Big]. \nonumber \\
\end{eqnarray}
Combining (\ref{classical1}) and (\ref{covaction}) we construct the covariant form of 
Maxwell action with the one-loop quantum correction in a four dimensional space time characterized by 
$\tilde{\mathcal{G}}_4$. In the next section we study the gauge field dynamics governed by the covariant quantum corrected 
action. In particular, with the boundary condition set at large $u$, 
we solve the equation of motion for gauge field. Furthermore, by utilizing this solution, we follow the
 Kubo's prescription to compute a frequency dependent conductivity.

\section{Computation of frequency dependent conductivity} \label{four}
We consider the covariant form total effective action constructed in four dimensional curved space time,
\begin{eqnarray}
  S_{total} = \frac{1}{2} \int d^4 x \frac{1}{g_s^{eff}} \Big[ \dot{\tilde{\Xi}}^2 \Psi_1 - \tilde{\Xi}'^2 \Psi_2 \Big],
\end{eqnarray}
where $\Psi_1$ and $\Psi_2$ are the following,
\begin{eqnarray}
\Psi_1 (u)&=&  \sqrt{-Det\mathcal{\tilde{G}}_4} \frac{1}{u^3f(1-\frac{e^2}{fu^3})}
\Big \{ 1 - g_s^{eff} \mathcal{C}(I_1)\Big \}, \nonumber \\
\Psi_2(u) &=&  \sqrt{-Det\mathcal{\tilde{G}}_4}\frac{f }{u^3 (\frac{1}{u^3}+f {X_4'}^2)} 
\Big \{ 1 - g_s^{eff} \mathcal{C} (I_2)\Big \}.
\end{eqnarray}
To write the total action consistently we have set $2 \pi l_s^2 = 1$. The relative sign between the classical and quantum part is 
still to be defined. By comparing the standard 
form $e^{iS_{\text{real}}}e^{-\Gamma} = e^{i(S_{\text{Real}} + i\Gamma)}$
 we conclude that the sign of the normalization constant is negative. For simplicity we set $\mathcal{C} = -1$. 
 Assuming the space time structure of the gauge field 
$\tilde{\Xi}(u, t) = \tilde{\Xi}(u) e^{-i \omega t}$ we get the equation of motion,
\begin{eqnarray}
\tilde{\Xi}''(u) \Psi_2(u) + \tilde{\Xi}'(u) \Psi_2'(u) + \omega^2 \Psi_1(u) \tilde{\Xi}(u) = 0.
\label{eomnew}
\end{eqnarray}
The solution of the equation of motion with appropriate boundary conditions contains both normalizable and non-normalizable modes. 
The associated Green's function $G_R$ is defined as the ratio of normalizable to non-normalizable modes. 
Using Kubo formula in linear response theory, the working formula for the conductivity can be constructed from the structure of $G_R$. 
\begin{eqnarray}
 \sigma = \frac{G_R}{i \omega}.
 \label{sigma}
\end{eqnarray}
Before obtaining a solution of (\ref{eomnew}) we make a suitable coordinate transformation. 
\begin{eqnarray}
 \mathcal{Y}^2 = 1-\frac{u}{u_0}.
\end{eqnarray}
In this coordinate system, the equation of motion takes the following form,
\begin{eqnarray}
 \Big[\frac{d^2\tilde{\Xi}}{d \mathcal{Y}^2} \frac{{(1-\mathcal{Y}^2)}^4}{4u_0^2\mathcal{Y}^2} - \frac{d \mathcal{Y}}{d\mathcal{Y}}
 \frac{(3\mathcal{Y}^2 +1){(1-\mathcal{Y}^2)}^3}{4u_0^2 \mathcal{Y}^3} \Big] \Psi_2 
 + \Big[ \frac{d\tilde{\Xi}}{d \mathcal{Y}}  \frac{d\Psi_2}{d \mathcal{Y}} {\frac{{(1-\mathcal{Y}^2 )}^4}{4u_0^2 \mathcal{Y}^2}}\Big] 
 \nonumber \\+ \omega^2 \tilde{\Xi} \Psi_1(u) =0.
\label{eomfull}
\end{eqnarray}

The range of $\mathcal{Y}$ is fixed as $[0,-1]$. 
In the nine dimensional world volume theory of the $D8/\overline{D8}$, the above choice of $\mathcal{Y}$ coordinate 
has a natural interpretation. The location where the $D8$ and $\overline{D8}$ meet is denoted by $\mathcal{Y} = 0$, whereas 
the intersection points between the $D8$ and $\overline{D8}$ with $D4$ are specified as $\mathcal{Y} = \pm 1$. For
convenient nomenclature we refer $\mathcal{Y} = 1$ as the ``boundary'' and $\mathcal{Y} =-1$ as the `` horizon''.
For all subsequent analysis we set $u_{T} = 1$. The equation of motion for $\tilde{\Xi}(u) $ near the ``horizon'' 
becomes,
\begin{eqnarray}
 (1+ \mathcal{Y})\tilde{\Xi}''( \mathcal{Y}) -\frac{1}{2} \tilde{\Xi}'( \mathcal{Y}) + 
\frac{2\omega^2}{u_0}\tilde{\Xi}(\mathcal{Y}) = 0.
\label{eom1}
\end{eqnarray}
While deriving the equation (\ref{eom1}) we have used the asymptotic form of $\Psi_1$ and $\Psi_2$.
\begin{eqnarray}
\Psi_1 &\approx& \frac{\sqrt{1-{\mathcal{Y}}^2}}{\sqrt{u_0}}\frac{1}{\hat{g}_s^{eff}}, \nonumber \\
\Psi_2 &\approx& \frac{u_0^{5/2}}{{(1-y^2)}^{5/2}}\frac{1}{{\hat{g}}_s^{eff}}.
 \label{psi1}
\end{eqnarray}

We aim to solve to the above equation of motion near ``horizon'' in leading order of $ \mathcal{Y}$. Considering the newly
defined argument $\mathcal{Z}$, satisfying
\begin{eqnarray}
 \Psi_2 \frac{d}{d\mathcal{Y}} = \frac{d}{d\mathcal{Z}},
\end{eqnarray}
we re-write the equation (\ref{eom1}) as follows
\begin{eqnarray}
 \frac{d^2\tilde{\Xi}}{d^2\mathcal{Z}}+\omega^2 \Psi_1\Psi_2 \tilde{\Xi} = 0.
\end{eqnarray}
Choosing a suitable ansatz leads to the following set of equations,
\begin{eqnarray}
\tilde{\Xi} &=& e^{-S(\mathcal{Z})}, \nonumber \\
\tilde{\Xi}'' &=& -S''(\mathcal{Z}) e^{-S(\mathcal{Z})}+{S'(\mathcal{Z})}^2 e^{-S(\mathcal{Z})}. 
\end{eqnarray}
Ignoring $S''(\mathcal{Z})$ for the leading order solution we get,
\begin{eqnarray}
S = \int^\mathcal{Z} \pm i\omega \sqrt{\Psi_1\Psi_2} d\mathcal{Z}'.
\end{eqnarray}

The form of $\Psi_1$ and $\Psi_2$ at $\mathcal{Y} = -1$ is easy to derive from (\ref{psi1}).
Therefore, by solving the equation (\ref{eom1}), we get the leading term in $\tilde{\Xi}$,
\begin{eqnarray}
 \tilde{\Xi} = e^{\pm i \frac{2\sqrt{2}\omega}{\sqrt{u_0}} \sqrt{1+\mathcal{Y}}}.
\end{eqnarray}
For computation of the conductivity usually one has to employ the ingoing boundary condition at some location in the 
space time where 
the flavor brane touches the real horizon of black hole background.   
In our analysis, neither the flavor branes touch the black hole 
horizon nor any induced horizon gets developed in the  world volume. Similar kind of situation has been discussed in \cite{KSSS}, 
where the background holographically corresponds to a confining phase of a strongly coupled system at finite temperature. 
Since they have studied a system with real Lagrangian, they make a choice of
a complex boundary condition at $\mathcal{Y} = -1$ to get a 
finite real part of the conductivity. However, in our 
case the Lagrangian is complex. So the most general choice is the standing wave ansatz, i.e, the real boundary condition,
\begin{eqnarray}
\tilde{\Xi} &=& c_1 e^{ i \frac{2\sqrt{2}\omega}{\sqrt{u_0}} \sqrt{1+\mathcal{Y}}} + 
c_1^* e^{ -i \frac{2\sqrt{2}\omega}{\sqrt{u_0}} \sqrt{1+\mathcal{Y}}}, \nonumber \\
&=& \cos[\frac{2\sqrt{2}\omega}{\sqrt{u_0}} \sqrt{1+\mathcal{Y}}+\phi],
\end{eqnarray}

where $c_1$ is an arbitrary constant and $\phi$ is the phase.
We make a choice at $\mathcal{Y}=-1+\epsilon$, $\phi= -\frac{2\sqrt{2}\omega}{\sqrt{u_0}} \sqrt{\epsilon} $.
As a result of this choice, we can define the boundary conditions,  
\begin{equation}
 \tilde{\Xi}(\mathcal{Y} = -1) \sim 1, ~~~~~~\tilde{\Xi}' (\mathcal{Y} = -1) \sim 0.
\label{bndcond}
\end{equation}
At the ``boundary'', $\mathcal{Y}=+1$ the equation of motion simplifies as,
\begin{eqnarray}
 (1-\mathcal{Y})\tilde{\Xi}'' +\frac{1}{2} \tilde{\Xi}' + \frac{2\omega^2}{u_0}\tilde{\Xi}= 0.
\end{eqnarray}
The series solution for $\tilde{\Xi}(\mathcal{Y})$  near $\mathcal{Y}=+1$ reads as,
\begin{eqnarray}
 \tilde{\Xi}(y) &=& M \Big[1+M_1 (1-y) + M_2 {(1-y)}^2 + .....\Big],  \nonumber \\
 && ~~~~~~~~~~~~~~~                            + N{(1-y)}^{3/2} \Big[1+N_1 (1-y) + N_2 {(1-y)}^2 + .....\Big],
\end{eqnarray}
where $M$ and $N$ are the parameters to be determined by boundary conditions and $M_i$ and $N_i$ depend on the 
parameters of the theory.
The Green's function of our interest is given by,
\begin{eqnarray}
 G_{R} \sim \frac{N}{M} = \frac{4}{3}\frac{\sqrt{1-\mathcal{Y}}}{\tilde{\Xi}(\mathcal{Y})} \frac{d^2\tilde{\Xi}}{d\mathcal{Y}^2}.
\label{Greenfn}
\end{eqnarray}

\begin{figure}[h]
 \centering
 \mbox{\subfigure[]{\includegraphics[width=7.0 cm]{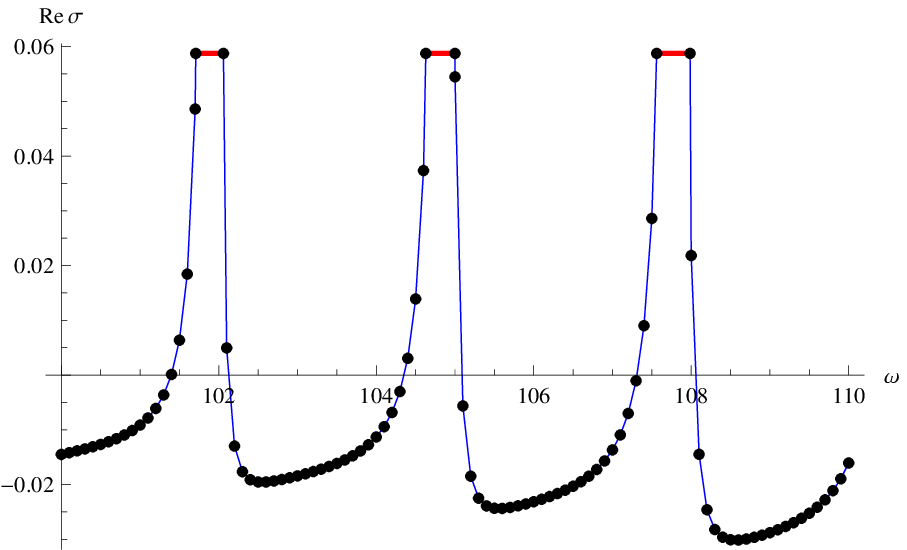}}
 \quad
\subfigure[]{\includegraphics[width=7.0 cm]{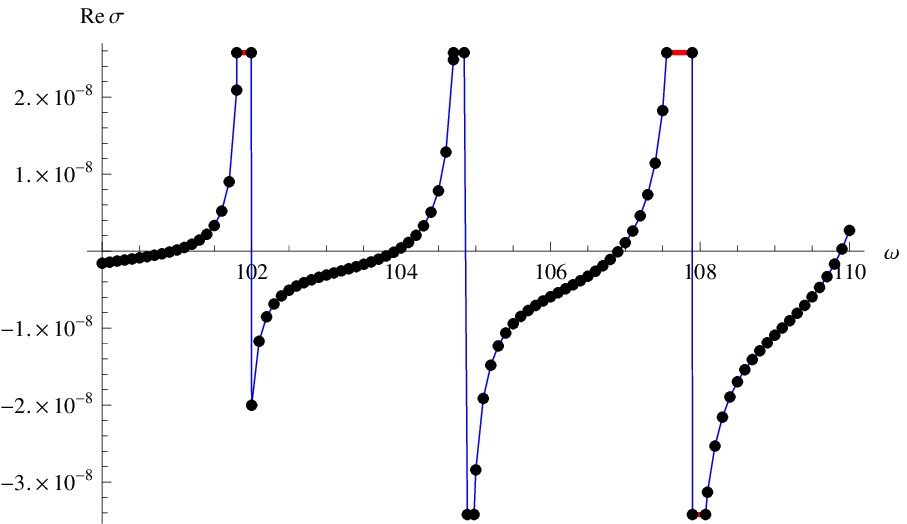} }

}
 \caption{Fig (a) and (b) depicts the real part of the conductivity for the set of parametric values 
($E = \frac{1}{10^2}$, $m = 5$) and ($E = \frac{1}{10^2}$, $m = 10$). Here we have chosen $u_0 = 10 $.}
 \label{condfig11}
 \end{figure}
 \begin{figure}[h]
 \centering
 \mbox{
\subfigure[]{\includegraphics[width=7.0 cm]{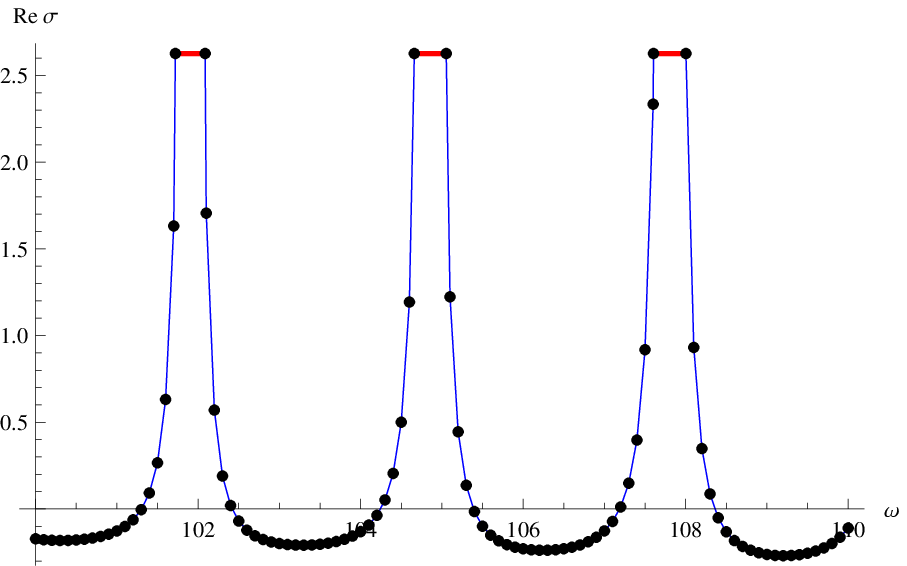} }
\quad
\subfigure[]{\includegraphics[width=7.0 cm]{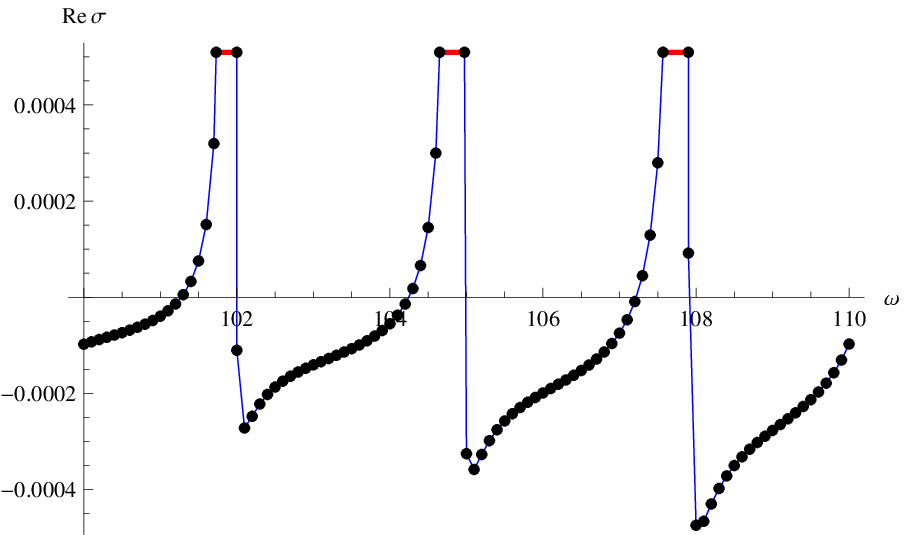} }
}
 \caption{Fig (a) and (b) shows the real part of the 
conductivity with the choices of the parameters are
 ($E = \frac{1}{50}$, $m = 5$) and ($E = \frac{1}{50}$, $m = 10$) respectively. Here we have chosen $u_0 = 10 $.}
 \label{condfig22}
 \end{figure}
 
 \begin{figure}[h]
 \centering
 \mbox{\subfigure[]{\includegraphics[width=7.2 cm]{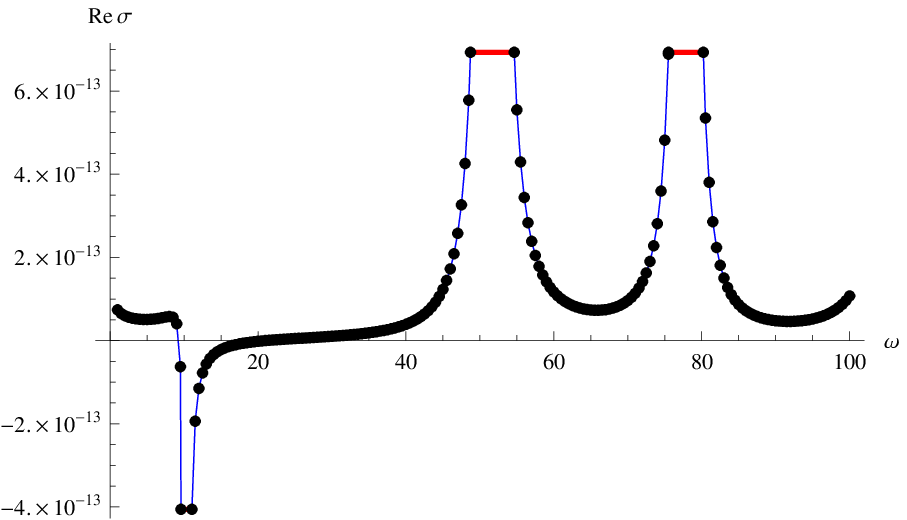}}
 \quad
\subfigure[]{\includegraphics[width=7.2 cm]{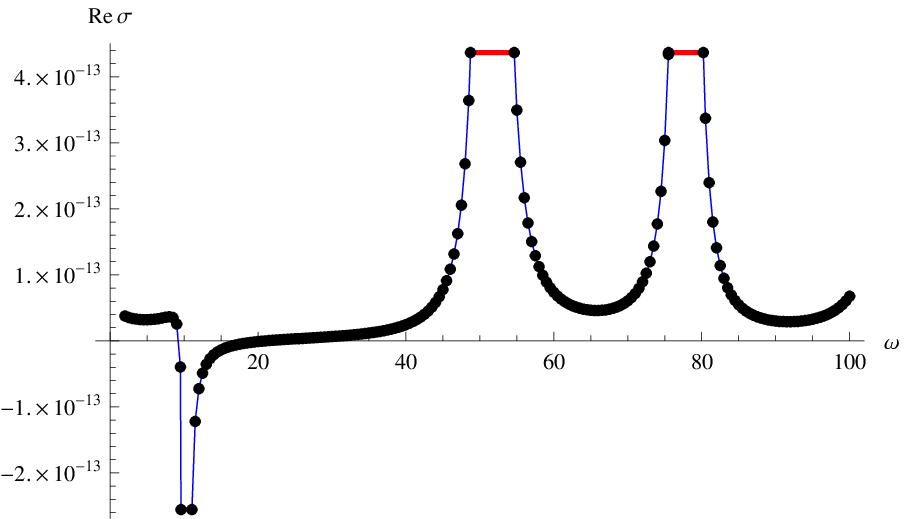} }}
 \caption{ In this figure we have plotted the real part of the $\sigma$ with respect to $\omega$ for 
 the set of parametric values 
 ($E = \frac{1}{30^3}$, $m = \frac{1}{100}$) and ($E = \frac{1}{2\times 30^3}$, $m = \frac{1}{100}$). Here we have chosen $u_0 = 1000$.}
 \label{condfig33}
 \end{figure}
 
  \begin{figure}[h]
 \centering
 \mbox{\subfigure[]{\includegraphics[width=4.6 cm]{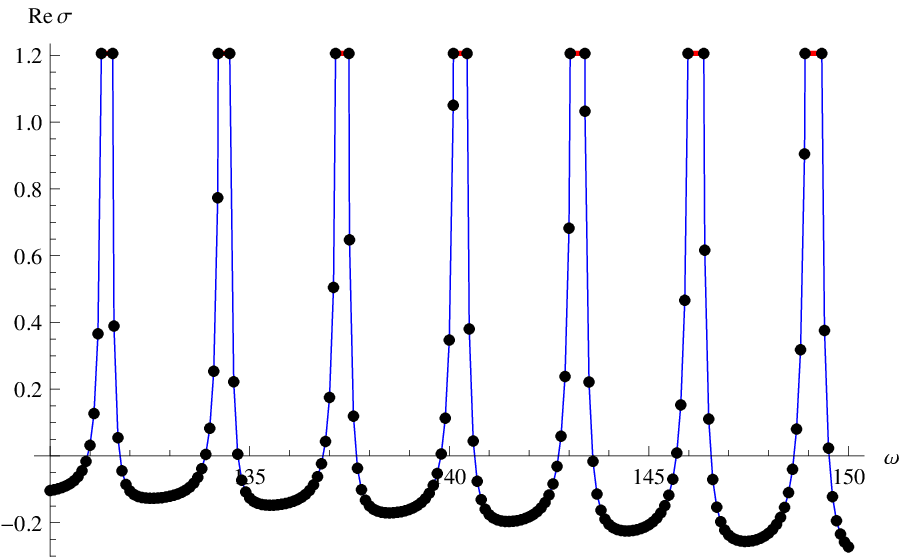}}
 \quad
\subfigure[]{\includegraphics[width=4.6 cm]{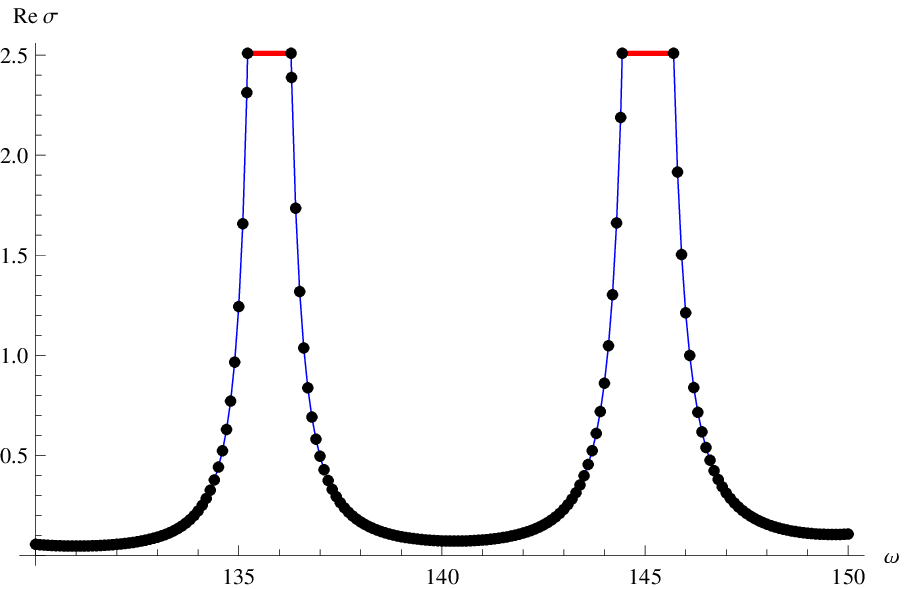} }
\quad
\subfigure[]{\includegraphics[width=4.6 cm]{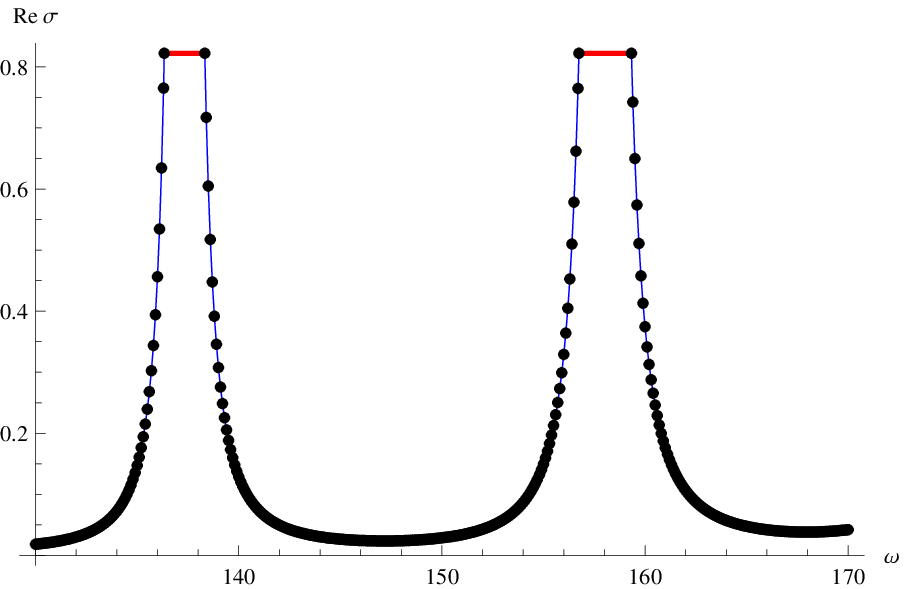} }

}
 \caption{Fig (a), (b) and (c) depict the variation of the separation between two consecutive peaks in Re $\sigma$ vs $\omega$ plot 
 for three different 
 values of $u_0 = 10, 100, 500$. The choice of others parameters are fixed as ($E = \frac{1}{2\times 30^3}$, $m = \frac{1}{100}$)}
 \label{condfig44}
 \end{figure}

Finally the conductivity is calculated by inserting the form of $G_R$ in (\ref{sigma}). Here we actually solve the equation (\ref{eomfull}) 
subject to the boundary conditions (\ref{bndcond}) using numerical methods. Then we apply the full solution together with (\ref{sigma}) and 
(\ref{Greenfn}) to compute the conductivity at the `` boundary'' $\mathcal{Y} = 1$. Finally, we plot the real part of the conductivity with respect to 
the frequency for both small and large $\lambda$. For example, figure (\ref{condfig11}) and  (\ref{condfig22}) 
show the variation of the real part of $\sigma$ with respect to frequency for $\lambda < 1$. We 
also show the same for  $\lambda > 1$ in figure  (\ref{condfig33}).
\pagebreak
From both cases it is evident that if we increase the mass, the conductivity is less. 
On the other hand conductivity enhances with the external electric field. For large and small 
$\lambda$, real part of the conductivity seems to diverge periodically with respect to $\omega$. The 
periodicity of this divergence almost remains constant as we vary the parameters $m$ and $E$. 
For $\lambda < 1$ we observe that the conductivity takes negative value with respect to the  higher values of frequency. 
For $\lambda > 1$, the region of negative conductivity lies within the lower range of $\omega$. 
For higher range of $\omega$ it is positive and always periodically divergent. In fig (\ref{condfig44}) 
we have plotted the conductivity for three different values of $u_0 = 10, 100, 500$ keeping the electric field 
$E = \frac{1}{2\times 30^3}$ and  $m = \frac{1}{100}$ fixed. We observe in this plot that for lower values of $u_0 = 10$ the real part 
of the $\sigma$ takes negative values. As we increase the $u_0 = 100, 500$ the conductivity is always positive. Most importantly, the 
frequency of the peaks in conductivity plots scales as $\sqrt{u_0}$. 
These peaks may arises from the poles of the Green's function and the corresponding frequency modes are well known as normal/quasi-normal modes. 
In the confined phase of Sakai-Sugimoto model, the normal modes determines the mass of the meson spectrum and the mass parameter turns out to 
be proportional to the Kaluza-Klein mass ($m_{KK}$) of the theory. 
Since $m_{KK}$ scales as the square root of $u_{KK}$, the period of  normal modes also scales as $\sqrt{u_{KK}}$. 
In our case, we have set the $u_{KK} =1$ and in the deconfined phase the role of $u_{KK}$ 
is played by the radius of horizon $u_{T}$ (the $u-x_4$ subspace is a topologically cylinder). Therefore the only relevant parameter 
to set the mass scale is $u_0$. As we already observe that the periodicity 
of the frequencies where the conductivity (evaluated numerically) seems to diverge  scales as $\sqrt{u_0}$. So it is tempting to conclude 
that these apparent numerical divergences are in fact  poles appearing in the Greens's function and actually characterize the quasinormal modes.

\pagebreak

\section{Conclusion} \label{five}

In this paper we have discussed the effect of pair production in modifying the conductivity 
in strong coupling situations. The strong coupling theory is  described holographically by a 
system of flavor and color branes. The flavor branes provide flavor quantum numbers to particles. 
The one loop Schwinger effect modifies the effective action for a photon on the flavor brane. Since the coupling constant $g^2_{YM}=g_s$ is small, this one loop calculation is reliable. The photon is dual to a conserved current in the boundary theory. The current current correlator in the boundary is modified. This modifies the conductivity of the boundary theory. The interesting point is that because the Schwinger effect introduces an imaginary part to the effective action, the modified conductivity in the boundary acquires a real part. Thus
an insulator starts to have a small real conductivity in the presence of an electric field. 
What is doubly interesting is that the differential conductivity for some range of parameters can be 
{\em negative}. This effect is known in condensed matter contexts in semiconductor diodes such as tunnel diodes and Gunn diodes although the mechanism of negative differential resistance is specific to the device. Here we have a strong 
coupling version of this. This is similar to the effect discussed in \cite{Nakamura-Ooguri,Nakamura1,Nakamura2}. 
The main difference is that in these papers  the effect was already there at the classical level on the flavor 
brane because of the presence of an open string metric horizon. In our case there is no such horizon and shows 
up only at one loop.

It would be interesting to identify experiments analogous to those suggested in \cite{ACG} where the Schwinger effect would be observable.

\section*{Acknowledgements}
The authors would like to acknowledge Ganapathy Baskaran, S. Kalyana Rama, Nemani V. Suryanarayana, 
Sudipta Mukherji, Amitav Virmani and Swarnendu Sarkar for various fruitful discussions. 
S.C would like to thank Sudipto Paul Chowdhury and Sunanda Patra for useful discussion.

\end{document}